\documentclass[aps, reprint, prl, twocolumn, superscriptaddress,amsmath,amssymb, longbibliography,noeprint]{revtex4-2}
\usepackage{times}
\usepackage{amsmath}
\usepackage{amssymb}
\usepackage{xfrac}
\usepackage{dsfont}
\usepackage{graphicx}
\usepackage{float}
\usepackage{braket}
\usepackage[normalem]{ulem}
\usepackage{bm}
\usepackage{bbm}
\usepackage{color}
\usepackage{xcolor}
\usepackage{notoccite}
\usepackage{titlesec}
\usepackage[colorlinks=true, urlcolor=blue, linkcolor=blue, citecolor=blue]{hyperref}

\titlespacing*{\section}{0pt}{2ex}{2ex}

\newcommand{\affcua}{MIT-Harvard Center for Ultracold Atoms, Research Laboratory of Electronics, and Department of Physics,
Massachusetts Institute of Technology, Cambridge, Massachusetts 02139, USA}

\newcommand{\Swinburne}{Optical Science Centre and ARC Centre of Excellence in Future Low-Energy Electronics Technologies, Swinburne University of Technology, Melbourne 3122, Australia}

\newcommand{\kB}{k_\textrm{B}}
\newcommand{\TF}{T_\textrm{F}}

\newcommand{\vF}{v_\textrm{F}}
\newcommand{\kv}{\mathbf{k}}

\begin{document}
\title{Thermography of the superfluid transition in a strongly interacting Fermi gas}

% Author List
\author{Zhenjie Yan}
\thanks{Present address: Department of Physics, University of California, Berkeley, California 94720.}
\affiliation{\affcua}
\author{Parth B. Patel}
\author{Biswaroop Mukherjee}
\affiliation{\affcua}
\author{Chris J. Vale}
\affiliation{\Swinburne}
\author{Richard J. Fletcher}
\author{Martin Zwierlein}
\affiliation{\affcua}

% Abstract
\begin{abstract}

Heat transport is a fundamental property of all physical systems and can serve as a fingerprint identifying different states of matter. In a normal liquid a hot spot diffuses while in a superfluid heat propagates as a wave called second sound. Despite its importance for understanding quantum materials, direct imaging of heat transport is challenging, and one usually resorts to detecting secondary effects, such as changes in density or pressure. Here we establish thermography of a strongly interacting atomic Fermi gas, a paradigmatic system whose properties relate to strongly correlated electrons, nuclear matter and neutron stars. Just as the color of a glowing metal reveals its temperature, the radiofrequency spectrum of the interacting Fermi gas provides spatially resolved thermometry with sub-nanokelvin resolution. The superfluid phase transition is directly observed as the sudden change from thermal diffusion to second sound propagation, and is accompanied by a peak in the second sound diffusivity. The method yields the full heat and density response of the strongly interacting Fermi gas, and therefore all defining properties of Landau's two-fluid hydrodynamics. Our measurements serve as a benchmark for theories of transport in strongly interacting fermionic matter.

\end{abstract}

\maketitle

%%%%%%%%%%%%%%%  Introduction %%%%%%%%%%%%%%%%%%%%%%%%

Heat transport is a ubiquitous phenomenon, at work in steam engines and the formation of stars, and dictates how energy, information and entropy flow in the system.
In conventional materials, heat, mass, and charge are all transported by the motion of (quasi-)particles, such as electrons in metals. This common origin of transport results e.g. in the Wiedemann-Franz law, relating thermal and electrical conductivity.
However, in strongly correlated systems, such as high-temperature superconductors~\cite{Lee2006}, neutron stars~\cite{Alford2018a}, and the quark-gluon plasma of the early universe~\cite{Adams2012}, the notion of a quasi-particle is poorly defined.
It is unknown whether there is a common relaxation rate for heat, density, and spin transport~\cite{Hartnoll2015} or if strong correlations separate these phenomena.
Understanding the flow of entropy is at the forefront of current research, with powerful techniques connecting thermal flow in quantum systems to gravitational duals~\cite{Adams2012,blake2017thermal}.
Directly measuring thermal transport, as distinct from mass or charge transport, is thus of great relevance to elucidate the origin of heat dissipation in strongly correlated matter.

Strongly interacting atomic Fermi gases near a Feshbach resonance provide an ideal platform for quantitative studies of fermion transport~\cite{Giorgini2008RMP,Ketterle2008a,Zwierlein2015c,Zwerger2016a,Krinner2017a}. As a result of scale invariance in resonant Fermi gases~\cite{Ho2004}, measurements constrain the equation of state and transport properties of other strongly interacting Fermi systems, including neutron matter at densities 25 orders of magnitude higher. The system features the largest superfluid transition temperature $T_c$, relative to its density, of all known fermionic systems~\cite{Ku2012}.

Here we introduce a novel thermography method to image heat in interacting quantum gases, whose working principle is general and may be applied to electronic systems as well. The method only requires a temperature-dependent spectral response that can be locally resolved. In the case of the Fermi gas studied here, the radio-frequency (rf) spectrum is temperature dependent ~\cite{Mukherjee2019,Yan2019}. We spatially resolve this spectral response, and directly measure heat transport in the strongly interacting Fermi gas.

Remarkably, via heat transport one can distinguish states of matter. In ordinary liquids, heat transport is purely diffusive, and governed by thermal conductivity. In contrast, in superfluids, heat propagates as a wave called second sound. The two-fluid model of superfluidity introduces a normal and superfluid component which can move in and out of phase~\cite{Tisza1938,Landau1941}. This gives rise to two distinct sound modes, first and second sound, corresponding to a density and an entropy wave~\cite{Khalatnikov1965}.
The speed of second sound $c_2$ is a direct measure of the superfluid fraction ${{\rho_S}/{\rho}}$~\cite{suppmat}. Its attenuation yields the second sound diffusivity $D_2$, which involves the thermal conductivity, bulk and shear viscosities~\cite{Khalatnikov1965,Hohenberg1965a}.
Consequently, we observe a dramatic change in thermal transport as the Fermi gas is cooled below its superfluid critical temperature $T_c$. Simultaneously recording the complete density and heat response of the system to a known external perturbation allows us to completely characterize the two-fluid hydrodynamics of the strongly interacting Fermi gas~\cite{Kadanoff1963,Hohenberg1965a}.

Previous studies of thermal transport in quantum gases relied on the weak coupling between the density and temperature of the gas~\cite{Sidorenkov2013b,Brantut2013a,Husmann2018,Hoffmann2021}. This allowed the observation of second sound in Bose~\cite{Christodoulou2021,Hilker2022} and Fermi gases~\cite{Sidorenkov2013b,Hoffmann2021,Li2022} but without directly measuring heat propagation.
By employing a homogeneous box potential formed by light sheets, we observe running and standing waves of second sound, demonstrating multiple reflections of entropy waves from the walls of the box.
Our thermography works across the superfluid transition, allowing the observation of a pronounced peak in thermal diffusion at $T_c$, characteristic of critical behavior expected near second order phase transitions.

%%%%%%%%%%%%%%%%%%%%%%%  Figure 1 %%%%%%%%%%%%%%%%%%%%%%%%%%%%%%%

\section{Spectral Thermometry}

\begin{figure}
\centering
\includegraphics[width=3.375in]{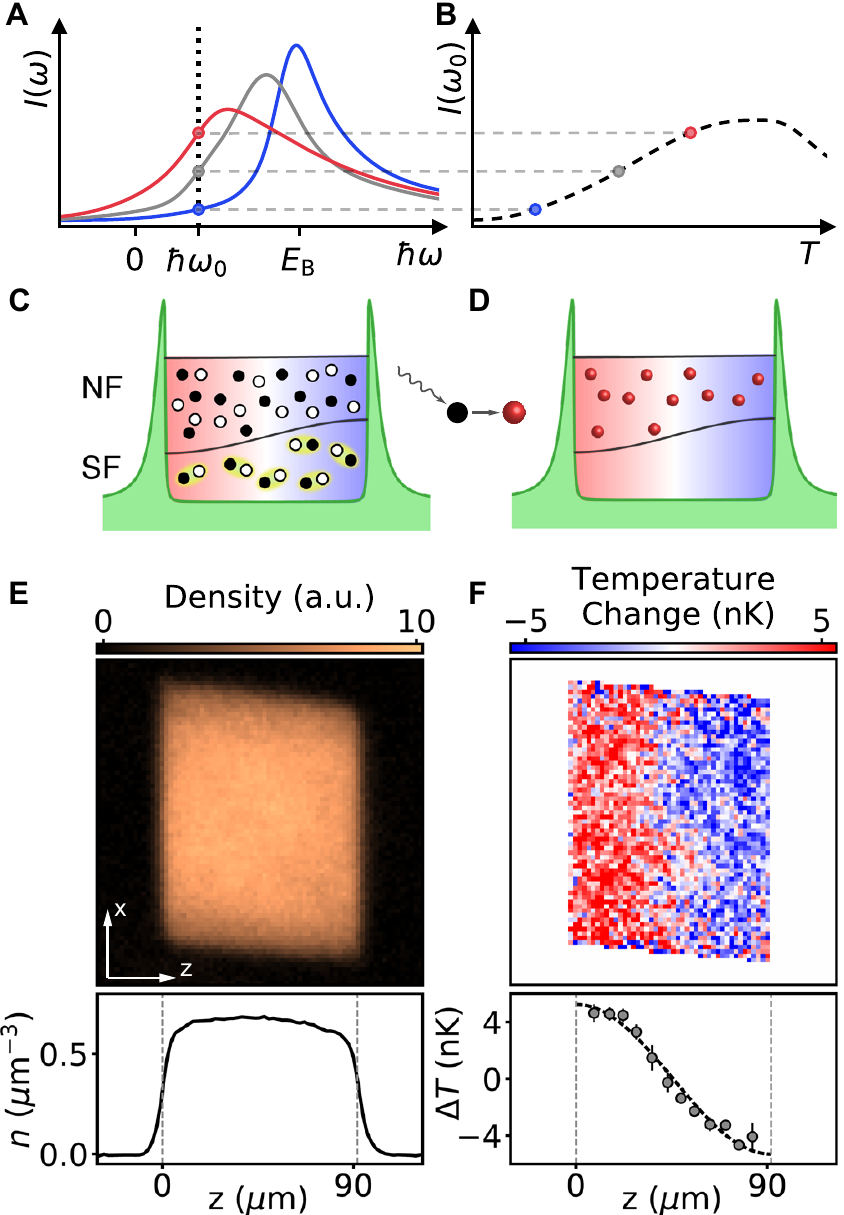}
\caption{\textbf{Direct local thermography using radio-frequency spectroscopy.} (\textbf{A}) A sketch of rf spectra at various temperatures for the unitary Fermi gas~\cite{Mukherjee2019}. Blue, grey, and red lines correspond to the rf response $I(\omega)$ at successively higher temperatures. (\textbf{B}) At fixed frequency $\omega_0$ on the flank of a spectrum (black dotted line in \textbf{A}), the rf response is sensitive to temperature, and serves as a local thermometer. (\textbf{C}) In a simplified picture, the superfluid component (SF) consists of fermion pairs, while the normal fluid (NF) is composed of broken pairs. (\textbf{D}) The unpaired atoms are transferred to a weakly interacting state by an rf pulse and subsequently imaged to determine the spatial distribution of the normal component density.
(\textbf{E}) and (\textbf{F}) In-situ observation of a second sound wave after resonant gradient excitation. Shown are the column density and local temperature, respectively, from simultaneous \textit{in situ} absorption images of unperturbed ($\ket{3}$) and rf-transferred ($\ket{2}$) atoms, with density $n$ and temperature variation $\Delta T$, averaged along the $x$-axis, shown below. The vertical dotted line marks the edge of the box potential (half-maximum of potential). The black dashed line in (F) is a fit to the fundamental eigenmode in the box. Second sound leaves a significant trace in the temperature, but not in the density.}
\label{fig:M1}
\end{figure}

The working principle of our method is sketched in Fig.\ref{fig:M1}A-D. In radiofrequency spectroscopy, interacting atoms are ejected from the many-body system into an initially unoccupied internal spin state~\cite{Gupta2003}. For interacting gases, such spectra depend on temperature: at high temperatures they approach the bare, unshifted response for an isolated atom, while at low temperatures they display interaction-induced shifts known as clock shifts. In the particular case of attractive two-component Fermi gases, at zero temperature the spectral peak is shifted by approximately the pairing energy $E_B$ of fermion pairs~\cite{Mukherjee2019}, while at non-zero temperature broken pairs contribute to the response at lower frequencies (Fig.~\ref{fig:M1}A). For a fixed detuning $\omega_0$ on the flank of the spectrum, the rf response is sensitive to changes in temperature (Fig.~\ref{fig:M1}B). As the rf response can be spatially resolved, this allows for a direct measurement of the local temperature from a single image of rf-transferred atoms.

%Figure 1

As a striking application of this method, we may detect second sound in the fermionic superfluid, which is a wave in the gas of excitations that close to $T_c$ consists predominantly of broken pairs (Fig.~\ref{fig:M1}C). A suitably detuned rf drive can transfer atoms from the gas of excitations, yielding a direct, local measure of heat (Fig.~\ref{fig:M1}D). We stress that the method does not depend on this simplified picture of broken pairs and only relies on the temperature-dependence of the rf spectrum. It therefore applies in a wide range of temperatures set by the magnitude of clock shifts, which for the unitary Fermi gas are on the scale of the Fermi temperature~\cite{Mukherjee2019}.

Our experiment starts with a uniform fermionic superfluid trapped in a cylindrical box potential, formed by an equal mixture of resonantly interacting fermions in the first ($\ket{1}$) and third ($\ket{3}$) hyperfine state of $^6$Li at a Feshbach resonance (magnetic field $690\ \mathrm{G}$)~\cite{Mukherjee2017b}. The density of $n_0=0.75\ \mathrm{\mu m ^{-3}}$ per spin state corresponds to a Fermi energy of $E_F=h \cdot 10.5\ \mathrm{kHz}$ and a Fermi temperature of $T_\mathrm{F} = E_F/k_B \simeq 500\ \mathrm{nK}$. To create temperature gradients in the superfluid gas, we resonantly excite a standing wave of second sound using an oscillating potential gradient along the axis of the cylindrical box~\cite{suppmat}.
Our thermography employs rf transfer of atoms from state $\ket{1}$ into the initially unoccupied state $\ket{f}\equiv\ket{2}$. Simultaneous {\it in situ} absorption images of atoms in states $\ket{2}$ and $\ket{3}$ yield the original gas density $n(\bold{r})$ (Fig.~\ref{fig:M1}E) as well as the density $n_\mathrm{f}(\bold{r})$ of rf transferred atoms, carrying the information on the local temperature (Fig.~\ref{fig:M1}F). 
The rf thermometer is calibrated on gases in thermal equilibrium by recording the dependence of $n_f$ on temperature, $\frac{\partial{n_f}}{\partial T}\Bigr|_n$, and density, $\frac{\partial{n_f}}{\partial n}\Bigr|_T$~\cite{suppmat}.
This method of calibrating spectral responses versus each thermodynamic variable while holding other parameters constant can be applied universally. More generally, all that is required for the observation of thermal transport is access to any local observable which is sensitive to temperature, meaning it can be achieved even without a calibrated thermometer.
Integrating the 2D temperature profile along the uniform radial direction yields a 1D temperature profile $\Delta T(z)$ with a precision of 500 picokelvin from a single image, as shown in Fig.~\ref{fig:M1}F.
The data reveals an essentially flat density in the presence of a $\sim 8\,\mathrm{nK}$ temperature difference across the box.

%%%%%%%%%%%%%%%%%%%%%%%%% Figure 2 %%%%%%%%%%%%%%%%%%%%%%%%%%%%%%%

%figure2
\begin{figure*}
\includegraphics[width=6.75in]{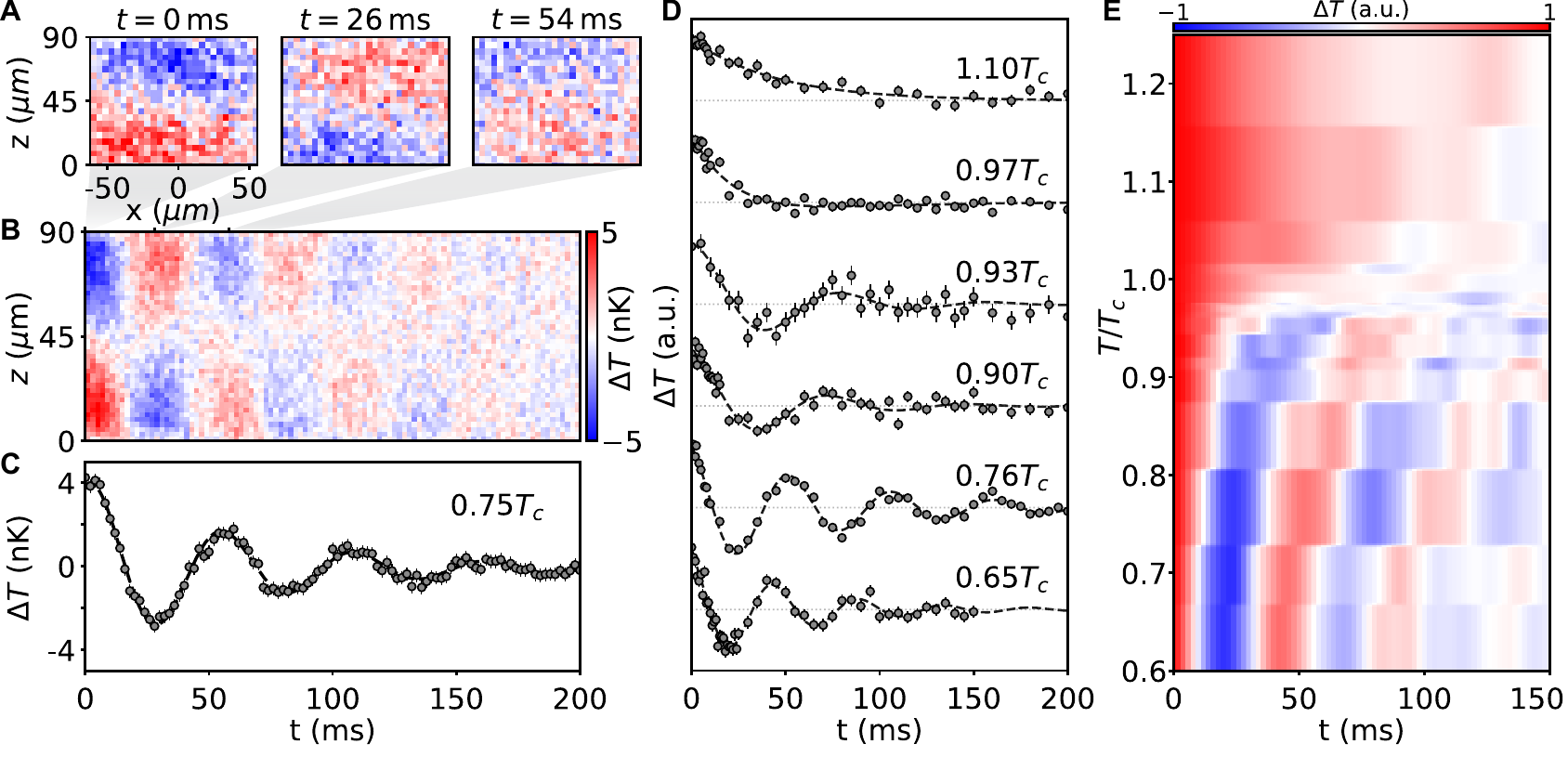}
\caption{\textbf{Direct observation of the superfluid transition from heat propagation in a strongly interacting Fermi gas.} (\textbf{A}) {\it in situ} thermographs at times $t{=}0,\,26$, and $54\,\mathrm{ms}$ after second-sound excitation, and
(\textbf{B}) time evolution of the axial temperature profiles, revealing the wave-like propagation of heat. (\textbf{C}) Amplitude of the first spatial Fourier mode of the temperature profiles $\Delta T (k_1,t)$ versus time (black circles). A fit to a damped sinusoid (dashed line) gives the speed and attenuation rate of second sound. In (A-C) the second sound is generated with resonant gradient excitation, and the gas temperature is $T=63\ \mathrm{nK}$ or $0.75\,T_c$.
(\textbf{D}) Time evolution of temperature amplitudes $\Delta T (k_1,t)$ (solid circles) and fits (dashed lines) at various gas temperatures. (\textbf{E}) 2D interpolation with gaussian smoothing of temperature amplitudes versus time across the superfluid transition. In (D-E) the second sound is generated with local heating, and the initial temperature variation for each time traces are normalized to be 1.
}
\label{fig:M2}
\end{figure*}

\section{Observation of heat propagation}
Armed with the ability to spatially resolve temperature in the strongly interacting Fermi gas, we directly observe one of the most striking manifestations of superfluidity, second sound, as the free back-and-forth sloshing of heat after resonant gradient excitation (see Fig.~\ref{fig:M2} A-C).
Fig.~\ref{fig:M2}A shows the measured rf transfer $\Delta n_\mathrm{f}(\bold{r},t)$ obtained at various times after second-sound generation. Fig.~\ref{fig:M2}B presents the time evolution of the 1D temperature profiles $\Delta T(z,t)$, and Fig.~\ref{fig:M2}C shows the corresponding evolution of the amplitude $\Delta T(k_1, t)$ of the first spatial Fourier mode supported by the box ($k_m= m \pi/L$), all clearly demonstrating the wave-like propagation of heat.
Here the absolute temperature of the gas in equilibrium, obtained from expansion~\cite{Yan2019}, is $T=63(2)\ \mathrm{nK}=0.125(5)\ T_\mathrm{F}$, or $T=0.75(3)\ T_c$ when compared with the superfluid transition temperature $T_c=0.167$ reported in Ref.~\cite{Ku2012}.
A damped sinusoidal fit to $\Delta T(k_1, t)$ yields a speed of second sound of $c_2 = \omega/k = 3.57(2) \mathrm{mm/s}$, corresponding to about a tenth of the Fermi velocity $c_2=0.092(2)\,\vF$. From the measured damping rate $\Gamma$ we obtain a diffusivity of second sound $D_2 = \Gamma/k^2 = 2.44(11) \hbar/m$. As was found for the diffusivities of spin~\cite{Sommer2011c,Luciuk2017}, momentum~\cite{Cao2011a} and first sound~\cite{Patel2020}, a natural scale for the diffusivity of second sound is Planck's constant, divided by the particle mass~\cite{Yan2019b,Li2022}. This scale directly emerges in a strongly interacting quantum fluid from a mean-free path of carriers of approximately one interparticle spacing $d$, and characteristic speeds given by Heisenberg's uncertainty $\hbar/m d$~\cite{Sommer2011c}.
Remarkably, a similar scale of diffusivity is also measured for second sound in the strongly interacting bosonic superfluid $^4$He~\cite{Hanson1954}, while the more weakly interacting, fermionic $^3$He in its superfluid $A_1$ and $B$ phases displays much larger values of many hundreds to thousand times $\hbar/m$~\cite{Vollhardt1990}.

Thermography provides an unprecedented view of the superfluid transition in the strongly interacting Fermi gas. Figs.~\ref{fig:M2}D-E show the transition from heat diffusion in the normal state to wave-like propagation of heat, second sound, in the superfluid. For these data, we created a local hot spot on one side of the box by locally applying an intensity-modulated optical grating~\cite{suppmat}. Modulation at $\sim 2\,\rm kHz$ efficiently creates high-frequency phonons that rapidly decay into heat~\cite{Patel2020,Kuhn2020}, creating a temperature profile with good overlap with the $m{=}1$ mode. The subsequent evolution of the temperature amplitude $\Delta T(k_1,T)$ displays a striking change in character from exponential decay above $T_c$ to the damped sinusoid of second sound below $T_c$.

%%%%%%%%%%%%%%%%%%%%%%%%%%%%%%%%%% Figure 3 %%%%%%%%%%%%%%%%%%%%%%%%%%%%

%Figure 3 sonogram and waterfall
\begin{figure*}
\includegraphics[width=6.75in]{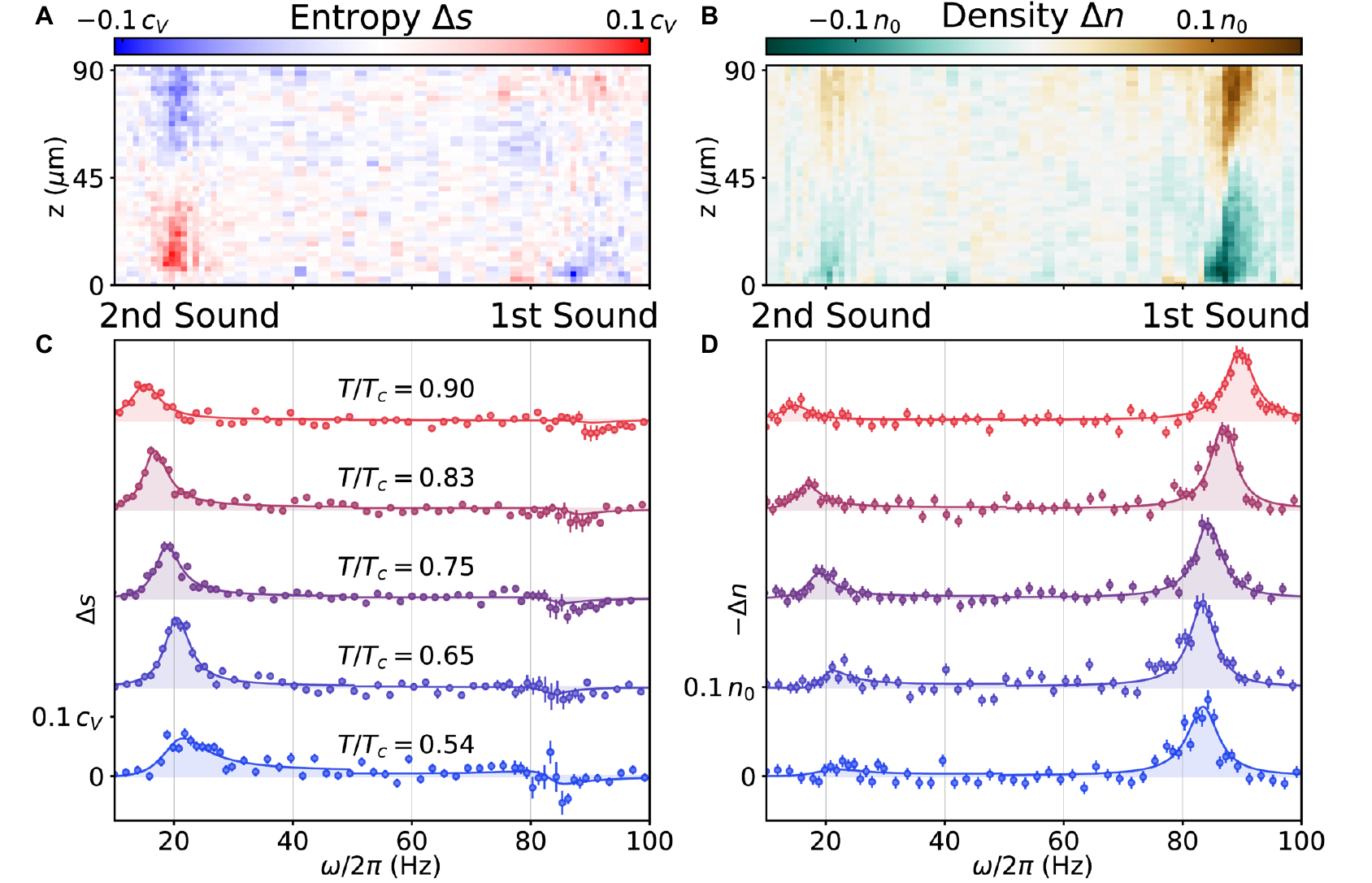}
\caption{\textbf{Steady state entropy and density response of the unitary Fermi superfluid.} Shown are the change in entropy per particle $\Delta s$ (\textbf{A}) and density $\Delta n$ (\textbf{B}) after excitation via an integer number of cycles of an oscillating axial potential gradient. For frequencies below 50 Hz, the drive duration is 5 cycles at an amplitude of $g=h \cdot 2.12\ \mathrm{Hz}/\mathrm{\mu m}$; for frequencies above 50Hz, we drive for 20 cycles at an oscillation amplitude of $g=h \cdot 0.85\ \mathrm{Hz}/\mathrm{\mu m}$. The gas temperature is $T/T_\mathrm{c}=0.75$. Amplitudes of the first spatial Fourier mode are shown in (\textbf{C}) and (\textbf{D}) for various temperatures in the superfluid phase. The solid lines are fits using the full entropy and density response function from two-fluid hydrodynamics~\cite{Hohenberg1965a,suppmat}.}
\label{fig:M3}
\end{figure*}

\section{Entropy and Density response functions}
The full linear response theory of two-fluid hydrodynamics for superfluids was provided over half a century ago by Hohenberg and Martin~\cite{Hohenberg1965a}. Under an external potential that acts on the density $n$ with wavevector $k$ and frequency $\omega$, systems respond through changes in their density $n$ as well as their temperature or equivalently entropy density $s$. Thermography enables us to obtain the corresponding response functions, not only $\chi_{n,n}(k,\omega)$ but also, for the first time in a quantum gas, $\chi_{s,n}(k,\omega)$. These encode all the thermodynamic and two-fluid hydrodynamic information of the unitary Fermi gas~\cite{Hohenberg1965a,Kadanoff1963,suppmat}.

To determine the linear response functions, we apply a potential gradient, oscillating at frequency $\omega$. The steady-state temperature change $\Delta T(k_1,\omega)$ and density change $\Delta n(k_1,\omega)$ measured after an integer number of oscillation cycles yield the respective out-of-phase response functions~\cite{Hohenberg1965a,Kadanoff1963}.
The change in entropy per particle $s$ is linked to the temperature and density variation by the equation of state. For our scale invariant, unitary Fermi gas, this connection is provided by the specific heat per particle $c_V$ at constant density ~\cite{Ho2004,Ku2012},
\begin{equation}
\Delta s=c_V \left( \frac{\Delta T}{T} -\frac{2}{3}\frac{\Delta n}{n_0} \right).
\label{dS_dTTF}
\end{equation}
Measurements of fractional temperature and density variations thus directly yield the entropy variation in units of $c_V$.
Fig.~\ref{fig:M3}A-B displays the entropy and density response of the superfluid in a frequency range that solely excites the lowest spatial mode ($m{=}1$), the sloshing mode. 
The density reveals a dominant peak attributed to first sound near $90\,\rm Hz$~\cite{Patel2020}, and a faint signature of second sound at $20\,\rm Hz$, expected in a gas of non-zero expansivity, where density and temperature are coupled. However, in the entropy channel, whose signal derives predominantly from the rf transfer~\cite{suppmat}, the second sound peak yields a large response. This directly demonstrates that second sound in the unitary Fermi gas is predominantly an entropy wave, while first sound is essentially isentropic.
This is similar to the case in superfluid $^4$He~\cite{Khalatnikov1965} but drastically different from the case in 2D and 3D Bose gases, where density and entropy are strongly coupled~\cite{Verney2015,Christodoulou2021,Hilker2022}.
In Fig.~\ref{fig:M3}C and D, we show the thermal evolution of the entropy and density responses in the first spatial Fourier mode, which serve as a direct measurement of the out-of-phase entropy-density $\mathrm{Im}\chi_{s,n}(k_1,\omega)$ and  density-density $\mathrm{Im}\chi_{n,n}(k_1,\omega)$ response functions~\cite{suppmat}.
The measured response functions completely encode all information about the two-fluid hydrodynamics in a unitary Fermi gas~\cite{Hohenberg1965a,Kadanoff1963,suppmat}.
The peak positions and widths give the speeds and diffusivities of first and second sound. While the speed of first sound is a direct measure of the energy of the gas~\cite{Patel2020}, the speed of second sound yields the superfluid density. The height of the second sound peak in the entropy-density response is given by the expansivity $\alpha_p$ of the gas, and the weight of the second sound vs the first sound response in the density-density response directly equals $\gamma-1$, where $\gamma=c_p/c_V$ is the ratio of heat capacities at constant pressure and density. These are related by the isothermal compressibility $\kappa_T$, heat capacity and temperature by $\gamma-1=T \alpha_p^2/(n \kappa_T c_V)$, and in particular for the unitary gas simply by $\gamma-1 = \frac{2}{3}\alpha_p T$.

%%%%%%%%%%%%%%%%%%%%%%%%%%%%%%%%%% Figure 4 %%%%%%%%%%%%%%%%%%%%%%%%%%%%%%%
\begin{figure}[!]
\includegraphics[width=3.375in]{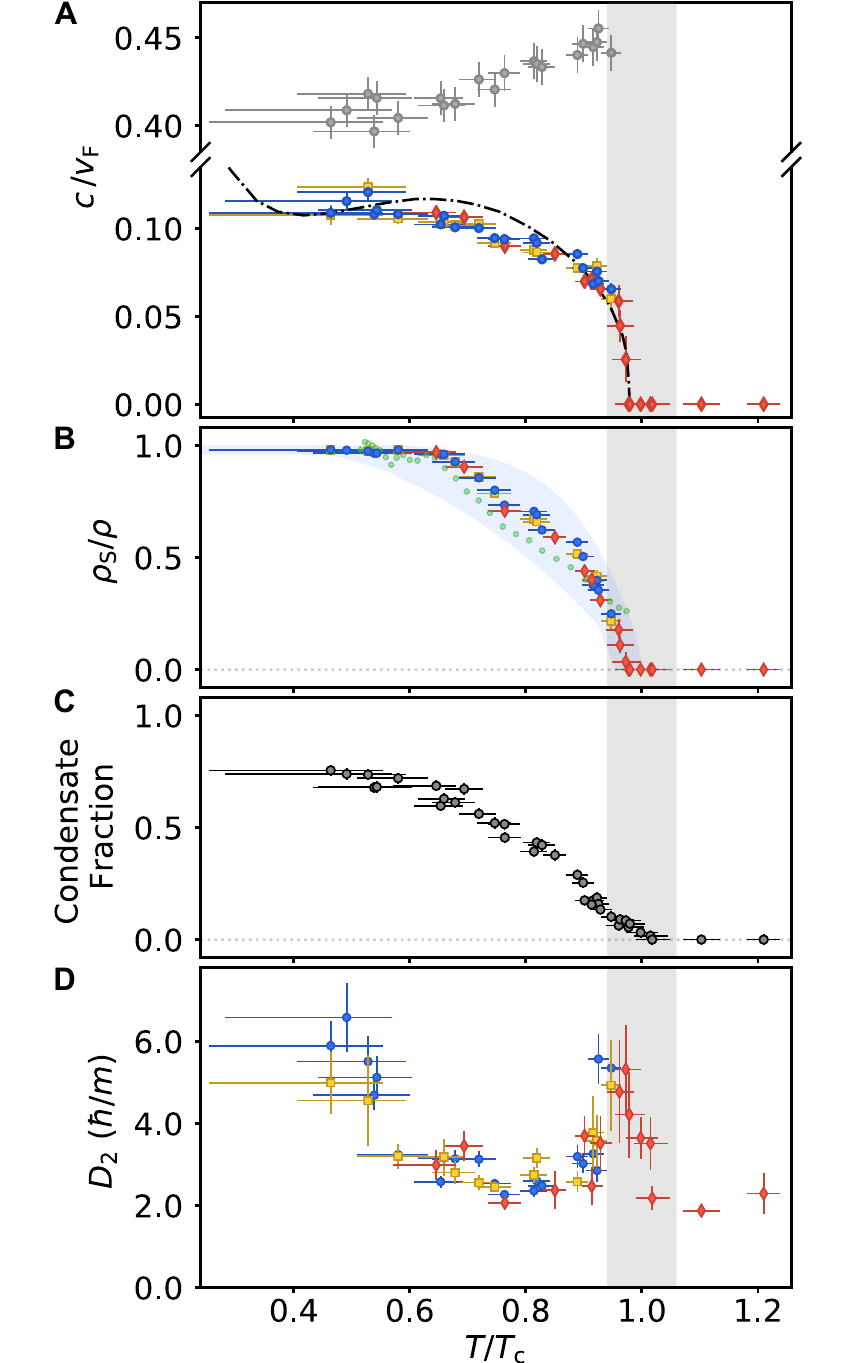}
\caption{\textbf{The speed and diffusivity of second sound.} (\textbf{A}) Speed of second sound, normalized by the Fermi velocity, as a function of temperature, determined by fitting the steady state response functions (blue circles), and the free evolution of second sound after resonant gradient excitation (yellow squares) or after local heating (red diamonds). The first sound speed measured from the response functions (gray circles) is also shown. 
Dot-dashed line: Nozi\`eres-Schmitt-Rink theory~\cite{Taylor2009}. 
(\textbf{B}) The superfluid fraction of the unitary Fermi gas obtained from the speed of second sound (symbols as in A). The blue shaded area indicates the uncertainty from the equation of state. Solid green circles: superfluid fraction obtained from quasi-1D experiments via the MIT equation of state~\cite{Sidorenkov2013b}. 
(\textbf{C}) Pair condensate fraction measured via the rapid ramp technique to detect fermion pair condensates~\cite{Mukherjee2019}.
(\textbf{D}) Second sound diffusivity obtained from various methods (symbols as in A). The vertical gray area indicates the uncertainty of critical temperature from Ref.~\cite{Ku2012}.
}
\label{fig:M4}
\end{figure}

\section{Heat transport across superfluid transition}
Fig.~\ref{fig:M4}A shows the speed of second sound, measured consistently using our three independent methods: free evolution after resonant excitation of the second sound mode (yellow squares), local heating (red diamonds) and steady state response functions (blue circles).
The superfluid fraction is obtained from $c_2$ and the previously measured equation of state~\cite{Ku2012,suppmat} and shown in Fig.~\ref{fig:M4}B.
The measurements show a qualitative agreement with Nozi\`eres and Schmitt-Rink theory~\cite{Taylor2009,Nozieres1985} (dot-dashed line), although their absolute value of $T_c$ differs from experiment.
Our superfluid fraction agrees well with the result reconstructed for the homogeneous case from the second sound measurement in a quasi-1D trapped gas in~\cite{Sidorenkov2013b}, which relied on the same equation of state of Ref.~\cite{Ku2012}.
With the local heating method (red diamonds) we are able to observe the continuous evolution of $c_2$ and $\rho_S$ from a finite value in the superfluid phase to zero in the normal phase. 
The phase transition temperature $T_c$ obtained from this measurement is consistent with the equilibrium thermodynamic measurement~\cite{Ku2012} (the vertical gray area) and the onset of pair condensation~\cite{Ketterle2008a,Mukherjee2019}, which we have measured here as well (Fig.~\ref{fig:M4}C).
As is expected, there is a clear quantitative difference between the superfluid fraction, which saturates to unity at temperatures $T \lesssim 0.1\,\TF$, and the pair condensate fraction, which remains less than $\sim 0.75$. The superfluid density quantifies the portion of the fluid that flows without friction. Formally it measures the rigidity against phase twists, while the condensate fraction is a measure for the number of fermion pairs at zero center of mass momentum.
In the zero-temperature limit, the entire system is superfluid, but only a fraction of fermion pairs are condensed, due to quantum depletion and Pauli blocking~\cite{Ketterle2008a,Giorgini2008RMP,Zwerger2016a}.

A further dramatic signature of the superfluid transition is seen in the temperature dependence of the second sound diffusivity $D_2$ in the superfluid state, and thermal diffusion in the normal state, shown in Fig.~\ref{fig:M4}D.
We observe a striking peak in this transport coefficient within a range $\Delta T \approx 0.1 T_c$ around the critical temperature of superfluidity, rising above a background minimum value of about $2\hbar/m$ up to nearly three times this value. This behavior echoes that found in liquid $^4$He~\cite{Hanson1954,Goldner1993} near its superfluid transition, associated with classical criticality. Indeed, the order parameters of both the Fermi superfluid and liquid helium belong to the same 3D XY static universality class, and also the same (model F in ref.~\cite{Hohenberg1977}) dynamic universality class, dictating a behavior $D_2 \propto |T_c - T|^{-\nu/2}$ near the transition, with critical exponent $\nu \approx 0.672$, as observed in $^4$He~\cite{Goldner1993}. Related critical behavior for the speed of second sound $c_2 \propto (T_c-T)^{\nu/2}$ and $\rho_S \propto (T_c-T)^{\nu}$ is qualitatively consistent with the steep slopes we observe close to $T_c$ in these quantities.
For the unitary Fermi gas, the width of the region governed by criticality is not precisely known, but estimated to be on the order of $T_c$~\cite{Taylor2009a,Debelhoir2016}.
A quantitative analysis of critical behavior, such as the measurement of critical exponents, is prevented by the residual inhomogeneity of the gas density, giving a variation of $\Delta(T/T_c) \sim 5 \times 10^{-3}$, and by the finite size of our system. Indeed, even for the lowest spatial mode $m=1$, second sound becomes overdamped ($\Gamma \gtrsim 2\omega$) within 3\% of $T_c$. 
At low temperatures $T/T_c < 0.6$, $D_2$ is again seen to rise significantly, which we attribute to the diverging mean-free path of phonons, the only remaining contribution at low temperatures once pair-breaking excitations are frozen out.

Above the transition temperature, the second sound mode evolves into a thermal diffusion mode whose diffusivity is directly given by thermal conductivity $\kappa$: $D_2 = \kappa/n c_P$~\cite{Kadanoff1963,Hohenberg1965a,Frank2020,Patel2022}.
We therefore find quantum limited thermal diffusion $\sim 2 \hbar/m$~\cite{Wang2022a}, similar to prior results for spin~\cite{Sommer2011c,Luciuk2017}, momentum~\cite{Cao2011a} and first sound diffusion~\cite{Patel2020} in the unitary gas.
However, the non-monotonous behavior of second sound diffusivity, with steep rise at low temperatures and around $T_c$ has not been observed in other transport coefficients.

The second sound diffusivity $D_2$ was independently measured via Bragg scattering~\cite{Li2022}, and a small rise in the second sound damping rate approaching $T_c$ was observed. However, a peak in $D_2$ near $T_c$ could not be resolved, presumably since Bragg scattering as a density probe becomes insensitive to heat propagation above $T_c$.
Away from $T_c$, the values for $D_2$ reported in ref~\cite{Li2022} were about half of what we observe.
Since the experiment in~\cite{Li2022} used a much higher wave vector and correspondingly more elevated frequencies, the gas may no longer have been hydrodynamic but instead entered the collisionless regime, similar to the behavior for high-momentum first sound in~\cite{Patel2020,Kuhn2020}. Assuming the hydrodynamic relation $\Gamma = D_2 k^2$ for such modes will yield too small a value for $D_2$.
In contrast, in the present work using thermography we verified hydrodynamic scaling by exciting also the second $m=2$ spatial mode supported by the box, finding within error bars identical values of $D_2$~\cite{suppmat}.

In the superfluid regime of the unitary Fermi gas, there are three contributions to second sound diffusion: thermal conductivity $\kappa$, shear viscosity $\eta$ as well as bulk viscosity $\zeta_3$ from normal-superfluid counterflow~\cite{Smith1989,Vollhardt1990}. While it is known that $\zeta_3 = 0$ for a pure phonon gas with linear dispersion~\cite{Escobedo2009}, in the range $T/T_c \gtrsim 0.5$ the normal fluid is dominated by pair breaking excitations. In this case, all three contributions are of similar importance~\cite{Smith1989,Vollhardt1990}. 
Assuming $\zeta_3=0$ in this regime, as was done in ~\cite{Li2022}, is not warranted, and obtaining viscosity and thermal conductivity from first and second sound diffusion alone is not possible.

%%%%%%%%%%%%Conclusion
\section{Outlook}
Direct measurement of heat transport has been a long-standing goal in quantum gas experiments. 
Thermography now opens the door to study a host of intriguing non-equilibrium phenomena, from non-linear heat waves to quench dynamics~\cite{Zwierlein2005,Dyke2021} and even far-from-equilibrium phenomena such as prethermal states~\cite{Gring2012,Eigen2018}.
Using tomographic imaging techniques~\cite{Ku2016}, the complete 3D spectral response can be measured, enabling the investigation of transverse entropy transport in anisotropic or inhomogeneous systems.
For thermodynamic systems with additional degrees of freedom beyond density and temperature, for example spin-imbalanced systems, additional independent measurements such as probes of the local spin polarization can be supplemented to fully determine thermodynamic response functions. The spectral response continues to serve as a channel highly sensitive to temperature.
Therefore, our spectroscopic thermometry method is general and can be applied to a wide variety of quantum gas platforms such as Bose gases, Bose-Fermi mixtures, impurity systems and Hubbard quantum simulators~\cite{Gross2017}, and more broadly to electronic or excitonic condensed matter systems.

%%%%%%%%%%%% Acknowledgements
\section{Acknowledgements}
We thank Boris Svistunov, Nikolay Prokof'ev, Wilhelm Zwerger and in particular the late Lev Pitaevskii for illuminating discussions.
This work was supported by the National Science Foundation (Center for Ultracold Atoms Awards No. PHY-1734011 and No. PHY-2012110), Air Force Office of Scientific Research (FA9550-16-1-0324 and MURI Quantum Phases of Matter FA9550-14-1-0035), Office of Naval Research (N00014-17-1-2257) and the Vannevar Bush Faculty Fellowship (ONR No. N00014-19-1-2631).

% \bibnotetext[suppmat]{See supplemental Information.}

\bibliography{main}

\onecolumngrid
\pagebreak

%% supplementary Information

%\part{}

\section[Supplementary Materials]{Supplementary Materials\\ Thermography of the superfluid transition in a strongly interacting Fermi gas}

\renewcommand{\thefigure}{S\arabic{figure}}
\setcounter{figure}{0}
\renewcommand{\theequation}{S\arabic{equation}}
\setcounter{equation}{0}

\section{Materials and Methods}

\begin{figure}[b]
\includegraphics[width=6.75in]{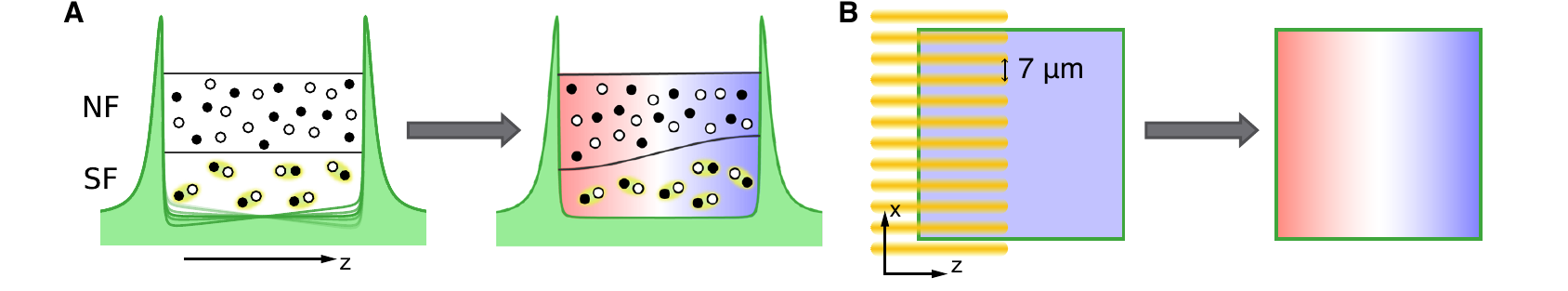}
\centering
\caption{\textbf{Two methods for creating second sound and heat gradients} (\textbf{A}) Second sound created by an oscillating potential gradient. The expansivity of the gas allows exciting temperature gradients through potential gradients.
(\textbf{B}) Creating temperature gradients using an intensity modulated light grating applied to one side of the box-trapped gas. Rapid modulations (at $2\,\rm kHz$) create short-wavelength phonons that rapidly decay into heat. In both \textbf{A} and \textbf{B} the blue to red color gradient represents an induced temperature variation from cold to hot.
}
\label{fig:S1}
\end{figure}

A strongly interacting Fermi gas of $^6$Li atoms is initially prepared in an equal mixture of the first and third lowest hyperfine states, ${\ket{1} \equiv \ket{m_J = -\frac{1}{2}, m_I=1}}$ and ${\ket{3} \equiv \ket{-\frac{1}{2}, -1}}$, at a magnetic field of $B=690 \,\rm G$, the location of the broad Feshbach resonance between these two states~\cite{Zurn2013}. The box potential is formed by walls of 532 nm light~\cite{Mukherjee2017b} and has cylindrical symmetry with a radius of $60\,\mathrm{\mu m}$ and a length of $90\,\mathrm{\mu m}$. 
The Fermi energy of the system is $E_F=10.5(3)\ h \cdot \mathrm{kHz}$ corresponding to a density of $n_0=0.75(3)\ \mathrm{\mu m ^{-3}}$  per spin state or a Fermi temperature of $T_\mathrm{F} \simeq 500\ \mathrm{nK}$.
Evaporation in the uniform trap leads to a temperature $T\lesssim 0.1\,\TF$.
An additional heating step via periodic modulation of the box potential is used to finely control the temperature of the cloud. 
To excite temperature gradients and second sound in the gas, we either apply an oscillating potential gradient, or alternatively we imprint an intensity-modulated light grating (Fig.~\ref{fig:S1}). 
The potential gradient is applied magnetically, by adding current on one of the magnetic field coils generating the Feshbach field. The amplitude of the magnetic field oscillation is less than 0.02 G, much less than the width of the Feshbach resonance ($\Delta B \approx 170\,\rm G$)~\cite{Zurn2013}, such that the modulation of atomic interactions is negligible. 
The oscillating gradient potential can generate a standing wave of first or second sound in the $m=1$ spatial mode, when the oscillation frequency is resonant.
The intensity-modulated grating potential serves as a local heat source in the atomic cloud. For this we employ 589 nm laser light to project a light-intensity grating onto the atoms. The grating spacing on the atomic cloud is $7\ \mathrm{\mu m}$. The intensity modulation is set to a frequency of 2 kHz, where phonons are efficiently created and rapidly decay into heat.
The amplitude of the grating modulation is kept at a low level so that the change in density and excitation of first sound is negligible.

To detect the temperature change in the system, a 0.5 ms rf pulse with a Rabi frequency of $\Omega_\mathrm{R}=2\pi\cdot1.9\ \mathrm{kHz}$ is used to transfer atoms in state $\ket{1}$ to the second lowest hyperfine state $\ket{f}\equiv\ket{2}=\ket{m_J = -\frac{1}{2}, m_I=0}$. 
For thermometry at gas temperatures below $T = 0.15 \TF \approx 0.9 T_c$, an rf detuning of $5$ kHz is used. For data obtained above $0.15\ \TF$, taken via the local heating method, a range of detunings from $2$ to $5$ kHz is used to improve temperature sensitivity.
The local densities $n_\mathrm{f}(\mathbf{r})$ and $n(\mathbf{r})$ of transferred atoms in state $\ket{f}$ and unaffected atoms in state $\ket{3}$ are imaged after the rf pulse.
The spatial resolution of rf thermometry is dominated by diffusion of the transferred $\ket{2}$ atoms through the background of $\ket{1}$-$\ket{3}$ atoms during the $0.5$ ms pulse time. The interaction between rf-transferred $\ket{2}$ atoms with atoms in states $\ket{1}$ and $\ket{3}$ is weakly repulsive with $k_F a_{23}=0.21$ and $k_F a_{12}=0.26$. The diffusivity of a $\ket{2}$ atom can be estimated with $D_{2i} \approx \frac{\hbar}{m}\frac{1}{(k_F a_{2i})^2}$, with $i\in \{1,2\}$, leading to a diffusion distance during the pulse time of about $10\,\rm \mu m$. This estimate agrees with direct measurements of this diffusion on a localized sample of $\ket{2}$ atoms, created in a $\ket{1}$-$\ket{3}$ background via localized Raman beams.

\section{RF Thermometry}

\begin{figure*}
\includegraphics[width=6.75in]{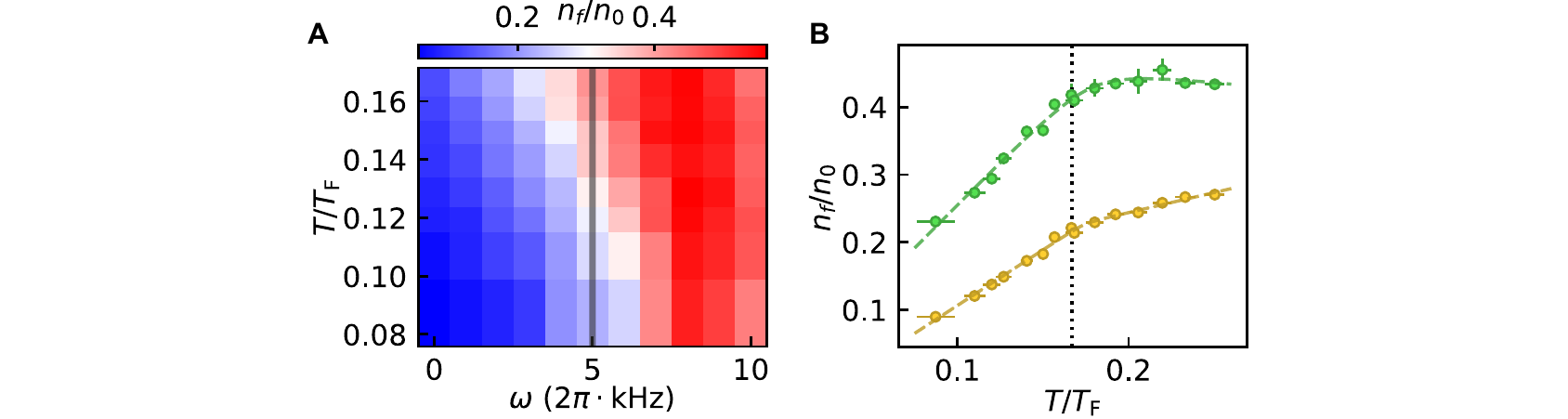}
\caption{\textbf{Calibration of the temperature dependence of the rf transfer at constant density $\frac{\partial{n_f}}{\partial T}\Bigr|_n$.} (\textbf{A}) Rf transfer fraction $n_f/n_0$ at various temperatures and detunings in the superfluid phase. In the regime below $T/\TF = 0.15$, we find a maximum linear gradient along the temperature axis at a detuning of $5\,\rm kHz$, marked by the gray vertical line.
(\textbf{B}) The temperature dependence of the rf transfer rate across the superfluid transition (vertical dotted line). The green (yellow) circles are the measured rf transfer fraction at a detuning of 5 kHz (2 kHz). 
The dashed lines are hyperbolas used to fit the rf transfer rate.
}
\label{fig:S2}
\end{figure*}

\begin{figure*}[b]
\includegraphics[width=6.75in]{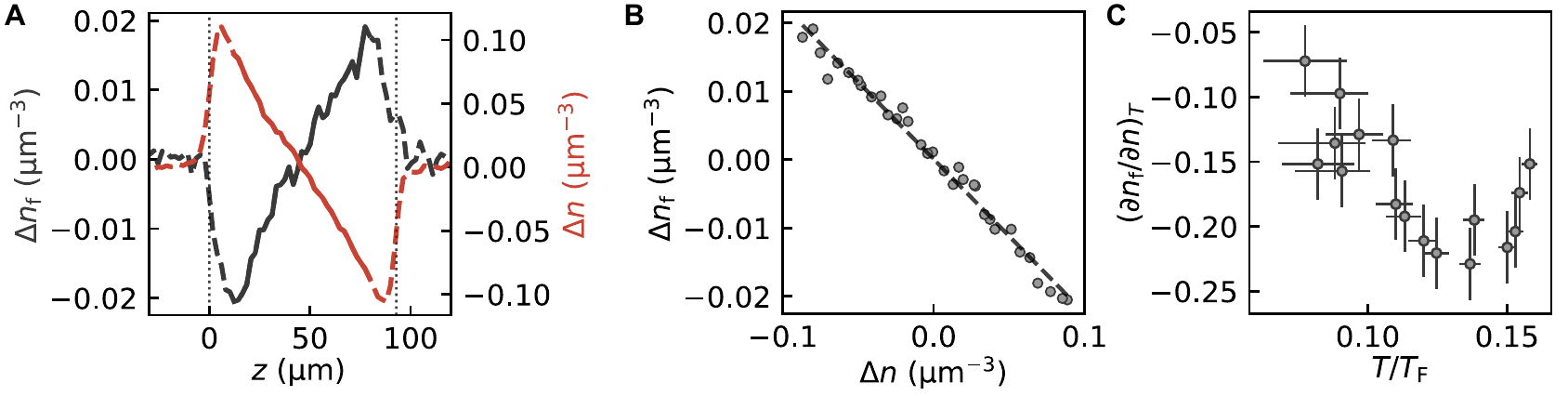}
\caption{\textbf{Calibration of the density dependence of the rf transfer at constant temperature $\frac{\partial{n_f}}{\partial n}\Bigr|_T$.} 
(\textbf{A}) Density change along the axial direction in the state $\ket{f}$ $\Delta n_\mathrm{f}(z)$ (blue) and state $\ket{3}$ $\Delta n(z)$ (red) after an adiabatic gradient ramp and the rf pulse. %
Only the data measured away from the edges of the cloud (marked by the vertical dotted lines) are used for the calibration (solid line).
(\textbf{B}) The linear response of rf transfer rate to the density change (black circles). $\frac{\partial{n_f}}{\partial n}\Bigr|_T$ is determined by a linear fit (dashed line).
(\textbf{C}) $\frac{\partial{n_f}}{\partial n}\Bigr|_T$ measured at various temperatures below $T_c$.
}
\label{fig:S3}
\end{figure*}

The sensitivity to temperature of the rf transfer is all that is needed to obtain a response from second sound and heat diffusion. The response would not need to be calibrated just to measure the speed and damping of second sound. However, by calibrating the response of the rf transfer to temperature, $\frac{\partial{n_f}}{\partial T}\Bigr|_n$ (rf-temperature response) and to density, $\frac{\partial{n_f}}{\partial n}\Bigr|_T$ (rf-density response), we can directly obtain the density-density and the density-temperature response functions of the unitary Fermi gas.
We calibrate the rf transfer at various temperatures and rf detunings, working at constant atom density (Fig.~\ref{fig:S2}A).
The temperature of the cloud is determined from the total energy measured through an isoenergetic expansion from the box potential into a harmonic trap~\cite{Yan2019} and converting this to temperature via the equation of state of the unitary Fermi gas from Ref.~\cite{Ku2012}.
For temperatures between $\sim 0.08$ and $0.15\,\TF$, a detuning of $\omega=2\pi \cdot 5\ \mathrm{kHz}$ is optimal, for which the temperature dependence of the transferred fraction $n_f/n_0$ is almost linear in temperature (Fig.~\ref{fig:S2}B). 
Above the transition temperature, the rf transfer at $5\ \mathrm{kHz}$ detuning saturates with increasing temperature. Therefore, when working close to and above $T_c$, we employ a smaller detuning, as in the measurement of the thermal evolution after local heating.
A hyperbola is used to fit the rf-temperature responses and extract the local slope.
To calibrate the response of the rf transfer to density at constant temperature, we adiabatically ramp up a linear gradient potential along the axial $z$-direction and measure the change in both $n(z)$ and $n_\mathrm{f}(z)$, as shown in Fig.~\ref{fig:S3}A. The potential ramp is slow enough so that the temperature stays constant across the cloud. The transferred atom number is seen to decrease with increasing initial density, as expected for a clock shift that increases with density.
A linear fit to the $n_\mathrm{f}(\Delta n)$ curve gives the rf-density response, as shown in Fig.~\ref{fig:S3}B. The temperature dependence of the rf-density response $\frac{\partial{n_f}}{\partial n}\Bigr|_T$ is shown in Fig.~\ref{fig:S3}C, displaying maximal sensitivity, about a quarter of the actual density variation, near $T/T_F = 0.14$.

\begin{figure*}
\includegraphics[width=6.75in]{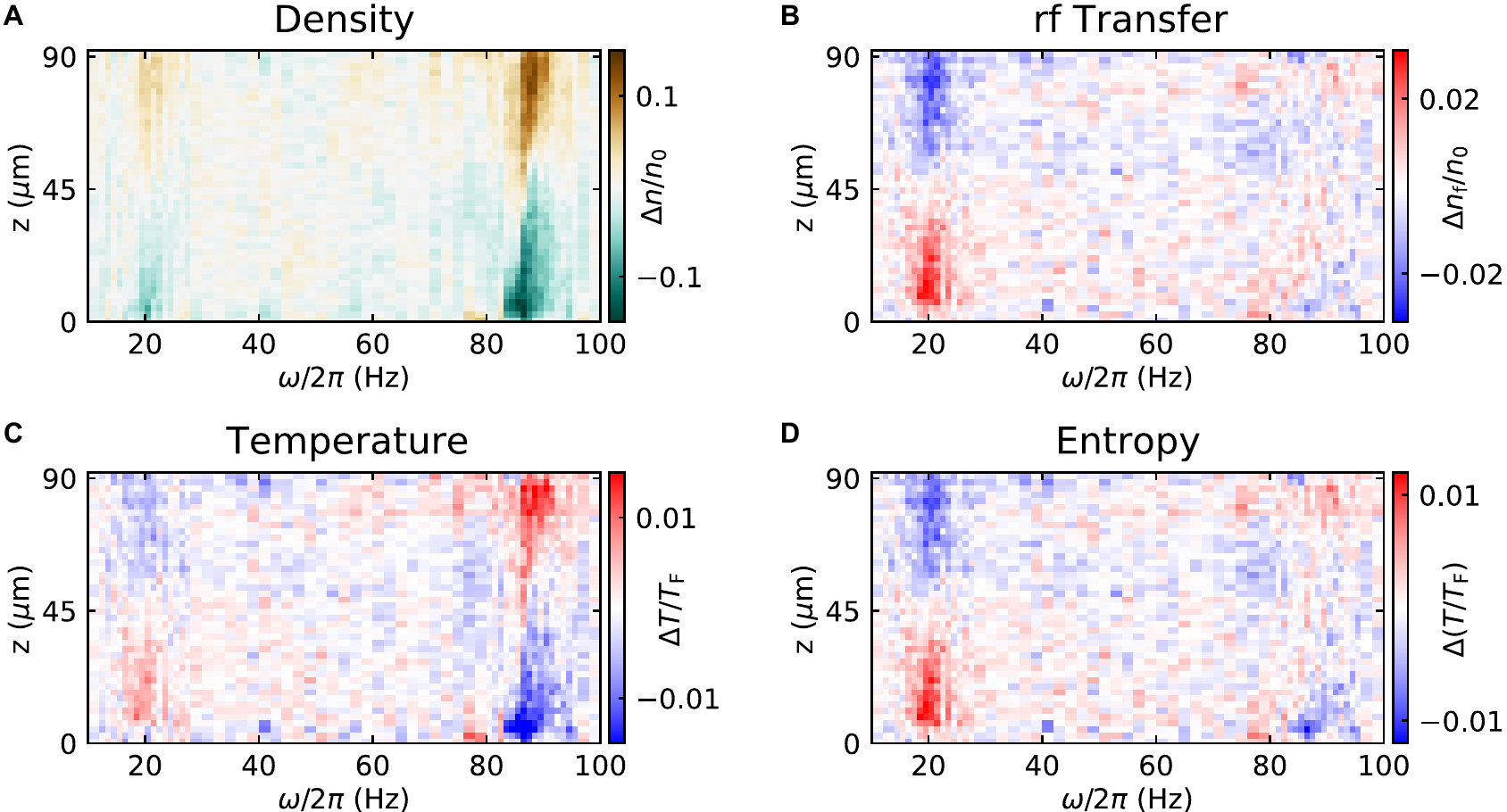}
\caption{\textbf{Steady state response of density $\Delta n$ (A), rf transfer $\Delta n_\mathrm{f}$ (B), temperature $\Delta T$ (C), and reduced temperature $\Delta (T/\TF)$ (D).} In a unitary Fermi gas, the reduced temperature corresponds to entropy per particle. The measurements shown here are done at a temperature of $T/\TF=0.125$.
}
\label{fig:S4}
\end{figure*}

Having calibrated the temperature and density dependence of the rf transfer, we obtain the local temperature change via the relation
\begin{equation}
\Delta T(\bold{r},t) = \frac{\partial T}{\partial n_\mathrm{f}}\Bigr|_n \left[\Delta n_\mathrm{f}(\bold{r},t) - \frac{\partial n_\mathrm{f}}{\partial n}\Bigr|_T \Delta n(\bold{r},t)\right].
\label{Temperautre}
\end{equation}
Our measurements capture another crucial thermodynamic quantity that is the change in the reduced temperature $\widetilde{T}=T/\TF$, given by the relation
\begin{equation}
\Delta \widetilde{T} (\bold{r},t)=\frac{\Delta T(\bold{r},t)}{T_\mathrm{F}}-\frac{2T}{3T_\mathrm{F} }\frac{\Delta n(\bold{r},t)}{n_0}.
\label{NormalizedT}
\end{equation}
In our scale invariant, unitary Fermi gas, the reduced temperature dictates the state of the system~\cite{Ho2004}. The entropy per particle $s = k_B f(\tilde{T})$ is a direct function of $\tilde{T}$, and the change $\Delta \tilde{T}$ directly yields the change in entropy via $\Delta s=c_V \Delta \tilde{T} / \tilde{T}$, since the specific heat $c_V = T\left.\partial s/\partial T\right|_n = k_B \tilde{T} f'(\tilde{T})$.
A measurement of the system's response in $\tilde{T}$ therefore directly yields the change in its entropy per particle, in units of the specific heat $c_V$.
Fig.~\ref{fig:S4} presents the individual responses of the density, rf transfer, temperature and reduced temperature after continuous oscillation of the potential gradient.
The density responds predominantly at the first sound resonance, with a faint response at the second sound resonance (Fig.~\ref{fig:S4}A). The rf transfer instead predominantly responds at the second sound resonance (Fig.~\ref{fig:S4}B).
Converting to absolute temperature changes $\Delta T$, we see that similar to the density response the strongest feature is from first sound. However, the response in reduced temperature $\Delta(T/T_F)$, which directly measures entropy changes, is strongest for the second sound resonance.
We have thus directly shown that second sound is an {\it entropy} wave (rather than a temperature wave) in a unitary Fermi superfluid, while first sound is an essentially isentropic density wave.
It is also evident that the rf transfer is a close proxy of the local entropy change. This is intuitively clear in the simplified picture of Fig.~\ref{fig:M1}, where the rf transfer predominantly excites the gas of excitations, e.g. pair breaking excitations, which directly determine the entropy of the gas.

\section{Testing hydrodynamic scaling of second sound damping}

The validity of hydrodynamics relies on local thermal relaxation being much faster than the frequency of a propagating disturbance. While damping rates of first and second sound are similar, second sound is much slower than first sound, and thus its frequency for a given spatial mode is much lower than that of the corresponding first sound. The criterion for full two-fluid hydrodynamicity is therefore much more stringent for heat transport than for density (mass) transport. The breakdown of hydrodynamics affects the expression for the sound damping rate $\Gamma$. While $\Gamma \propto k^2$ in the hydrodynamic regime, so that $\Gamma/k^2$ can be interpreted as a diffusivity, the damping rate rather scales linearly with $k$ in the collisionless regime. The crossover was experimentally observed in ~\cite{Patel2020} for first sound to occur at frequencies $\omega \lesssim 0.2\, mc_1^2 /\hbar$ at temperatures below $T_c$, where $c_1$ is the speed of first sound. For our system, for first sound, only the eight lowest box modes ($m\lesssim 8$) are safely in the hydrodynamic regime. For second sound, therefore, we should expect only the lowest two spatial modes ($m\lesssim2$) to display hydrodynamic scaling of their damping rates.

\begin{figure*}[!ht]
\includegraphics[width=6.75in]{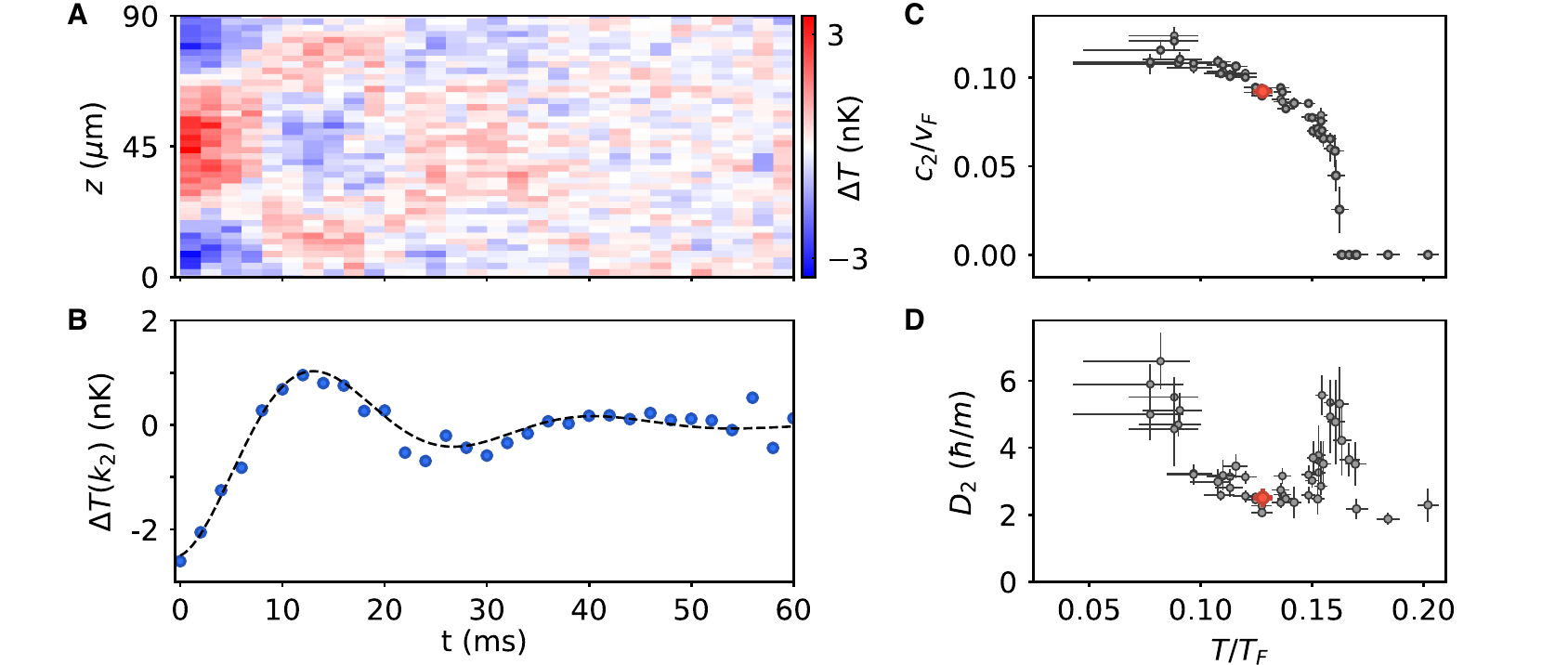}
\caption{\textbf{Testing hydrodynamic scaling with the $\mathbf{m=2}$ spatial mode of second sound.} (\textbf{A}) Axial temperature profile $\Delta T(z,t)$ versus time after resonant excitation of the second spatial mode $m=2$ supported by the box. (\textbf{B}) Fourier amplitude of the $m=2$ spatial mode of the temperature profile, $\Delta T(k_2,t)$ (blue circles). The dashed line is a fit with Eq.~\ref{eq:evolution}. The speed (\textbf{C}) and diffusivity (\textbf{D}) obtained from the evolution of the $m=2$ mode (red circle) is compared with the values measured for the $m=1$ sloshing mode (black circles), demonstrating linear dispersion of second sound and hydrodynamic scaling of damping rates. The data shown here for $m=2$ mode are taken with free evolution after resonant excitation.
}
\label{fig:S5}
\end{figure*}

To test hydrodynamic scaling, we were able to directly measure the frequency and damping rate of second sound in the $m=2$ mode. To excite this mode we imprint a potential using a 1064 nm laser beam focused at the center of the cloud, with a beam waist of $\sim 30\mathrm{\mu m}$, about one third of the box length. The intensity of the beam is modulated at the resonant frequency of the $m=2$ second sound mode, setting up a standing wave of second sound in $m=2$. Fig.~\ref{fig:S5}A shows the subsequent free time evolution of the temperature profile $\Delta T(z,t)$ after the end of the laser modulation. The corresponding Fourier amplitude $\Delta T(k_2,t)$ at $k_2=2\pi/L$ is shown in Fig.~\ref{fig:S5}B. As it should be for linear dispersion, the frequency of the $m=2$ is twice as high as for the corresponding $m=1$ mode (compare to Fig.~\ref{fig:M2}), i.e. the obtained speed of second sound is identical. The damping rate $\Gamma$ is found to be four times as high for $m=2$ than for $m=1$, as expected for hydrodynamic scaling. Fig.~\ref{fig:S5}C and D present the $m=2$ data points for speed and diffusivity together with the results for $m=1$. This consistency confirms that the observed damping rate $\Gamma$ for second sound is in the hydrodynamic regime for $m=1$ and $m=2$ and allows to uniquely define a second sound diffusivity $D_2= \Gamma/k^2$.

\section{Response functions}
The complete linear hydrodynamics of the two-fluid model of superfluids can be described via response functions as shown by Hohenberg and Martin~\cite{Hohenberg1965a}. 
The measured changes in density $\Delta n$, entropy $\Delta s$ and reduced temperature $\Delta (T/\TF)$ under a periodically modulated potential drive $U$ are associated with the density-density response function $\chi_{n,n}(\mathbf{k},\omega)$, the entropy-density response function $\chi_{s,n}(\mathbf{k},\omega)$, and the $\widetilde{T}$-density response function $\chi_{\tilde{T},n}(\mathbf{k},\omega)$:
\begin{align}
{\Delta} n (\kv,\omega) =& - \mathrm{Im}\chi_{n,n}(\mathbf{k},\omega)\cdot U(\kv,\omega), 
\label{res:E1A}\\
{\Delta} s (\kv,\omega) =& - {\mathrm{Im}\chi_{s,n}(\mathbf{k},\omega)} \cdot U(\kv,\omega), \label{res:E1B}\\
{\Delta} \widetilde{T} (\kv,\omega) =& - \mathrm{Im}\chi_{\tilde{T},n}(\mathbf{k},\omega)\cdot U(\kv,\omega) \label{res:E1C}
\end{align}
Since we measure the state of the system right after an integer number of modulation cycles, the relations only involve the imaginary part of the response functions, the out-of-phase response.
For a unitary Fermi gas, $\chi_{s,n}(\mathbf{k},\omega)$ and $\chi_{\tilde{T},n}(\mathbf{k},\omega)$ are simply related to each other via
\begin{equation}
    \chi_{\widetilde{T},n}(\mathbf{k},\omega)=\frac{\widetilde{T}}{c_V}\chi_{s,n}(\mathbf{k},\omega)
    \label{res:E1D}
\end{equation}
A spatial Fourier transform of the measured density or rf transfer responses is used to quantify the amplitude of density or reduced temperature changes at a certain wavenumber $k$:
\begin{equation}
{\Delta} n (k) = \frac{2}{L} \int_0^L \mathrm{d}z\ \Delta n(z) \cos(k z).
\label{Fourier}
\end{equation}
A linear gradient potential with slope $g$ only has a strong spatial Fourier component at the first fundamental mode of the box: $U(k_1=\pi/L)={4 g L}/{\pi^2}$. Hence, we mainly focus on the first Fourier mode of the responses in this work.

The general form of the density-density, entropy-density, and $\tilde{T}$-density response functions are~\cite{Hohenberg1965a}:
\begin{eqnarray}
\chi_{n,n}(\mathbf{k},\omega) &=& \frac{n_0 k^2}{m} \frac{-\omega^2+\gamma c_{20}^2 k^2-i \Gamma_{n,n} k^2\omega}{(\omega^2-{c_1}^2k^2+i D_1 k^2 \omega)(\omega^2-{c_2}^2k^2+i D_2 k^2 \omega)}, 
\label{res:E2A}\\
\chi_{s,n}(\mathbf{k},\omega) &=& k^2 \frac{-\alpha_p {c_{10}}^2{c_{20}}^2 k^2+i \Gamma_{s,n} k^2\omega}{(\omega^2-{c_1}^2k^2+i D_1 k^2 \omega)(\omega^2-{c_2}^2k^2+i D_2 k^2 \omega)},
\label{res:E2B}\\
\chi_{\widetilde{T},n}(\mathbf{k},\omega) &=& \frac{k_B T}{c_V E_F} k^2 \frac{-\alpha_p {c_{10}}^2{c_{20}}^2 k^2+i \Gamma_{s,n} k^2\omega}{(\omega^2-{c_1}^2k^2+i D_1 k^2 \omega)(\omega^2-{c_2}^2k^2+i D_2 k^2 \omega)}.
\label{res:E2C}
\end{eqnarray}
Here, $\gamma=c_p/c_V$ is the isentropic expansion coefficient. $\gamma-1$ controls the relative weight of the second to the first sound peak in the density response.
$\alpha_p=\frac{1}{V}{\frac{\partial V}{\partial T}}\bigr|_p$ is the thermal expansivity, and it controls the strength of the second sound peak in the entropy-density response. $\alpha_P$ also represents the static susceptibility of the change of entropy to an applied potential $U$. Indeed, one has the Maxwell relation giving the change in entropy per particle upon a change in chemical potential ${\rm d}\mu = -{\rm d}U$: 
$\left.\frac{\partial s}{\partial \mu}\right|_{T} = -\frac{1}{V}\left.\frac{\partial V}{\partial T}\right|_p = - \alpha_p$.
We note that at unitarity, $\alpha_P = \frac{3}{2}\frac{1}{T}(\gamma-1)$ is directly proportional to $\gamma-1$.

All response functions share the same denominator and poles.
The shared denominator is the determinant of the linear hydrodynamic equations, and the poles are located at
\begin{equation}
    \omega_{i,\pm}=\pm\sqrt{{c_{i}}^2k^2-({D_i} k^2/2)^2}-i{D_i} k^2/2,\ \ (i=1\ \text{or}\ 2).
\end{equation}
$c_{10}=\sqrt{(\partial p/ \partial{\rho})_S}$ and $c_{20}=\sqrt{\frac{s^2}{m}\frac{\rho_S}{\rho_N}\frac{\partial T}{\partial s}\bigr|_p}$ are the uncoupled speeds of first and second sound, and their values are given by the equation of state and for $c_2$ also the ratio between the densities of the superfluid component ($\rho_S$) to the normal component ($\rho_N$). The two sound modes are coupled by the isentropic expansion factor $\gamma$:
\begin{eqnarray}
{c_1}^2+{c_2}^2&=&{c_{10}}^2+\gamma {c_{20}}^2, \\
{c_1}{c_2} &=& {c_{10}} {c_{20}}.
\label{res:E3}
\end{eqnarray}
The diffusivities of the two sound modes $D_1$ and $D_2$ are both linear combinations of thermal conductivity $\kappa$, shear viscosity $\eta$, and bulk viscosities $\zeta_3$ in the unitary Fermi gas~\cite{Khalatnikov1965,Hohenberg1965a}.
Introducing $D_\kappa = \frac{\kappa}{\rho c_P}$, $D_\eta = \frac{4}{3}\frac{\eta}{\rho}$ and $D_\zeta = \rho \zeta_3$, as well as $D_{20} = D_\kappa + \frac{\rho_S}{\rho_N}(D_\eta+D_\zeta)$, one has the relations
\begin{align}
    D_1 + D_2 = D_\eta + D_{20} + (\gamma-1)D_\kappa\\
    c_1^2 D_2 + c_2^2 D_1 = c_{10}^2 D_{20} + c_{20}^2 \gamma D_\eta (1-2a)
\end{align}
with $a = \frac{1}{\rho s}\left.\frac{\partial p}{\partial T}\right|_n = \sqrt{\frac{\rho_S}{\rho_N}\frac{c_{10}^2}{c_{20}^2}\frac{\gamma-1}{\gamma^2}}$. For the unitary gas, this simplifies to $a = \frac{2}{3}\frac{c_V}{s}$.
Solving for $D_1$ and $D_2$, one obtains
\begin{align}
    D_1 = &\frac{c_1^2 (\frac{\rho_S}{\rho_N} +1)-c_{10}^2  \frac{\rho_S}{\rho_N} +(2a-1) c_{20}^2 \gamma}{c_1^2-c_2^2} D_\eta+\frac{\gamma c_1^2 - c_{10}^2}{c_1^2-c_2^2} D_\kappa+\frac{ (c_1^2 - c_{10}^2) \frac{\rho_S}{\rho_N} }{c_1^2-c_2^2} D_\zeta, \label{res:D1} \\
    D_2 = &\frac{c_{10}^2  \frac{\rho_S}{\rho_N} -c_2^2 \frac{\rho_S}{\rho_N}-(2a-1) c_{20}^2 \gamma}{c_1^2-c_2^2} D_\eta+\frac{c_{10}^2-\gamma c_2^2}{c_1^2-c_2^2} D_\kappa + \frac{(c_{10}^2-c_2^2) \frac{\rho_S}{\rho_N} }{c_1^2-c_2^2} D_\zeta.\label{res:D2}
\end{align}
A measurement of $D_1$ and $D_2$ alone does not allow to obtain the three individual diffusivities $D_\eta$, $D_\kappa$ and $D_\zeta$. However, the terms
$\Gamma_{n,n}$ and $\Gamma_{s,n}$, which govern the asymmetry of the sound peaks in the imaginary part of the response functions, are also functions of $\kappa$, $\eta$, and $\zeta_3$:
\begin{align}
    \Gamma_{n,n} & =\gamma D_\kappa  + \frac{\rho_S}{\rho_N} (D_\eta+D_\zeta) , \label{res:Gamma_nn} \\
    \Gamma_{s,n} & = \left(\frac{\rho_S}{\rho_N}D_\eta  + a\gamma D_\kappa \right) \frac{s}{m}.\label{res:Gamma_sn}
\end{align}
We see that the response functions enable to determine all thermodynamic and linear transport parameters of the unitary Fermi gas. In practice, determination of $D_\eta$, $D_\kappa$ and $D_\zeta$ separately involves taking the difference of noisy experimental data, leading to significant errors. Direct measurements of these quantities, e.g. by measuring the damping of shear flow to determine $\eta$ alone are necessary. 

\begin{figure*}
\includegraphics[width=6.75in]{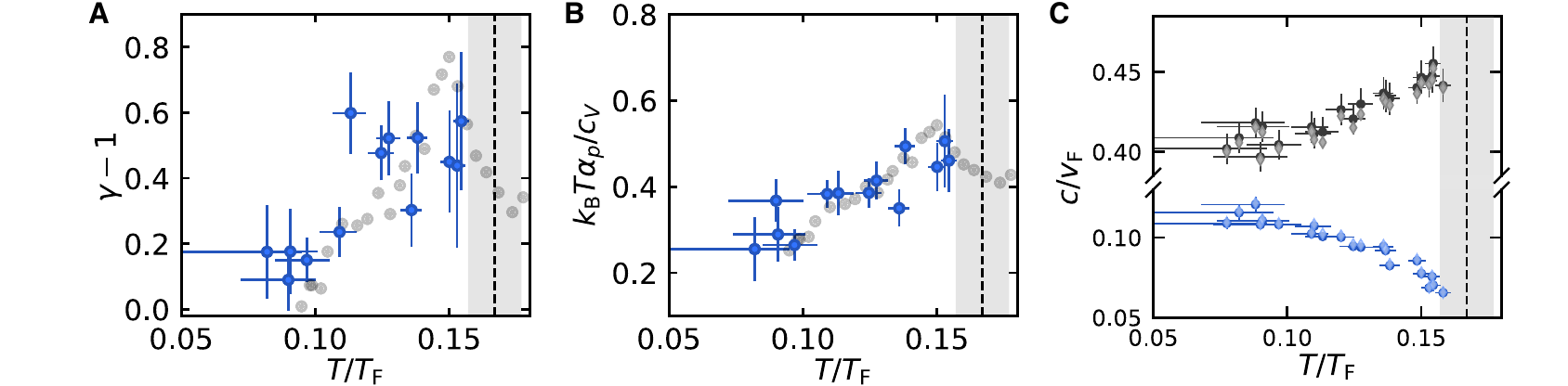}
\caption{\textbf{Thermodynamics from measured steady state response functions.} 
\textbf{(A)} The isentropic expansion coefficient $\gamma$, obtained from the measured density response (blue circles), compared to the value obtained from the previously measured equation of state~\cite{Ku2012} (gray circles).
\textbf{(B)} The amplitude of the normalized temperature response (blue circles), compared to the prediction $k_B T \alpha_p/c_V$ (gray circles) from the equation of state~\cite{Ku2012}.
\textbf{(C)} Coupled and uncoupled speeds of sound. The black circles (diamonds) and blue circles (diamonds) are the coupled (uncoupled) speeds of first and second sound, $c_1$ ($c_{10}$) and $c_2$ ($c_{20}$).
The vertical dashed line and shaded area indicates the phase transition temperature and its uncertainty, from~\cite{Ku2012}.
}
\label{fig:S6}
\end{figure*}

\begin{figure*}
\includegraphics[width=6.75in]{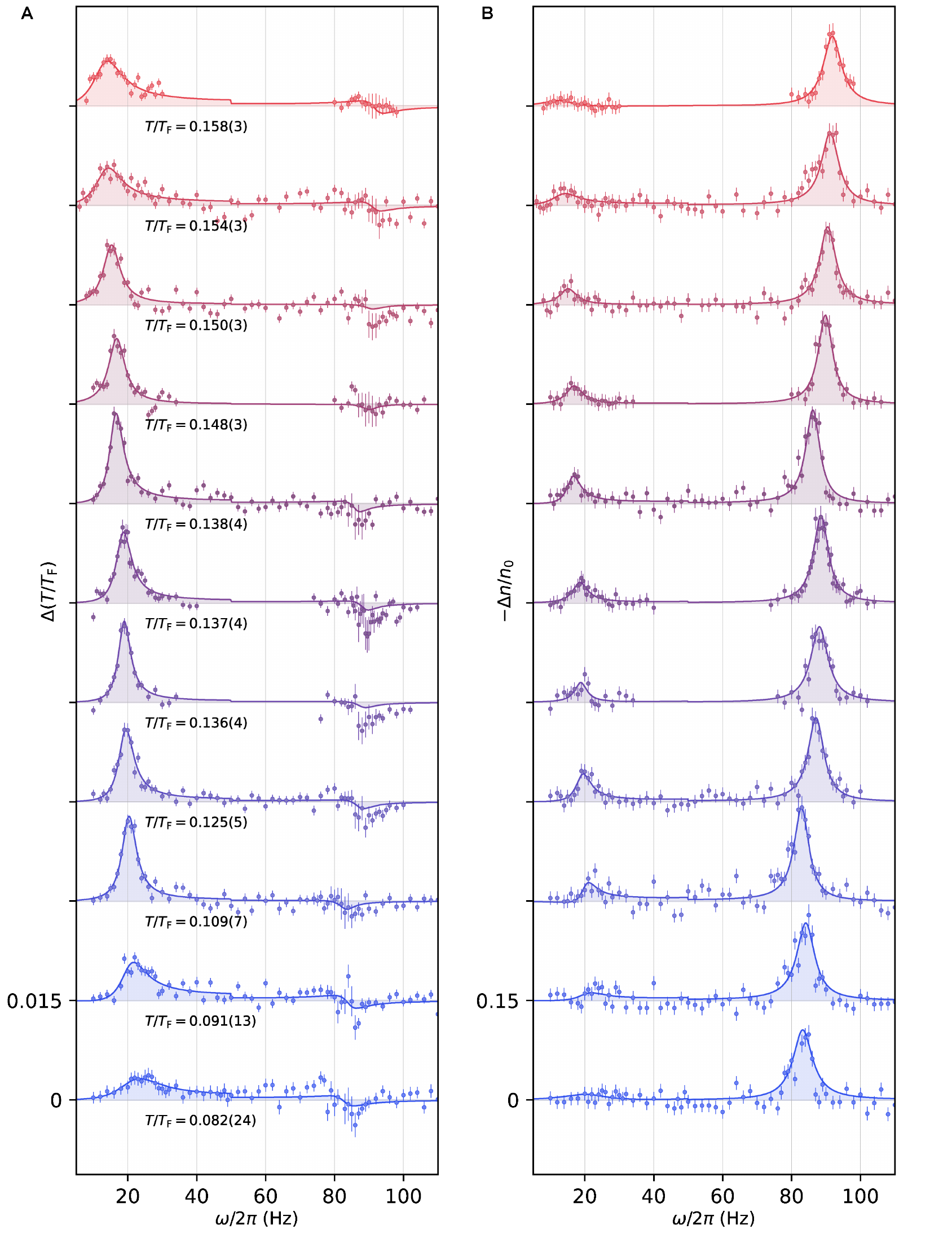}
\caption{\textbf{Steady state entropy and density responses under continuous modulation of a potential gradient.} (\textbf{A} and \textbf{B}) The normalized temperature (A) and density (B) response amplitude is obtained at wavenumber $k_1=\pi/L$. An oscillation amplitude of $g=h \cdot 2.12 \ (0.85)\ \mathrm{Hz}/\mathrm{\mu m}$ and 5 (20) shaking cycles is used for oscillation frequencies below (above) 50 Hz. The solid lines are fits using Eq.~\ref{res:E2A} and \ref{res:E2C}.
}
\label{fig:S7}
\end{figure*}

\begin{figure*}
\includegraphics[width=6.75in]{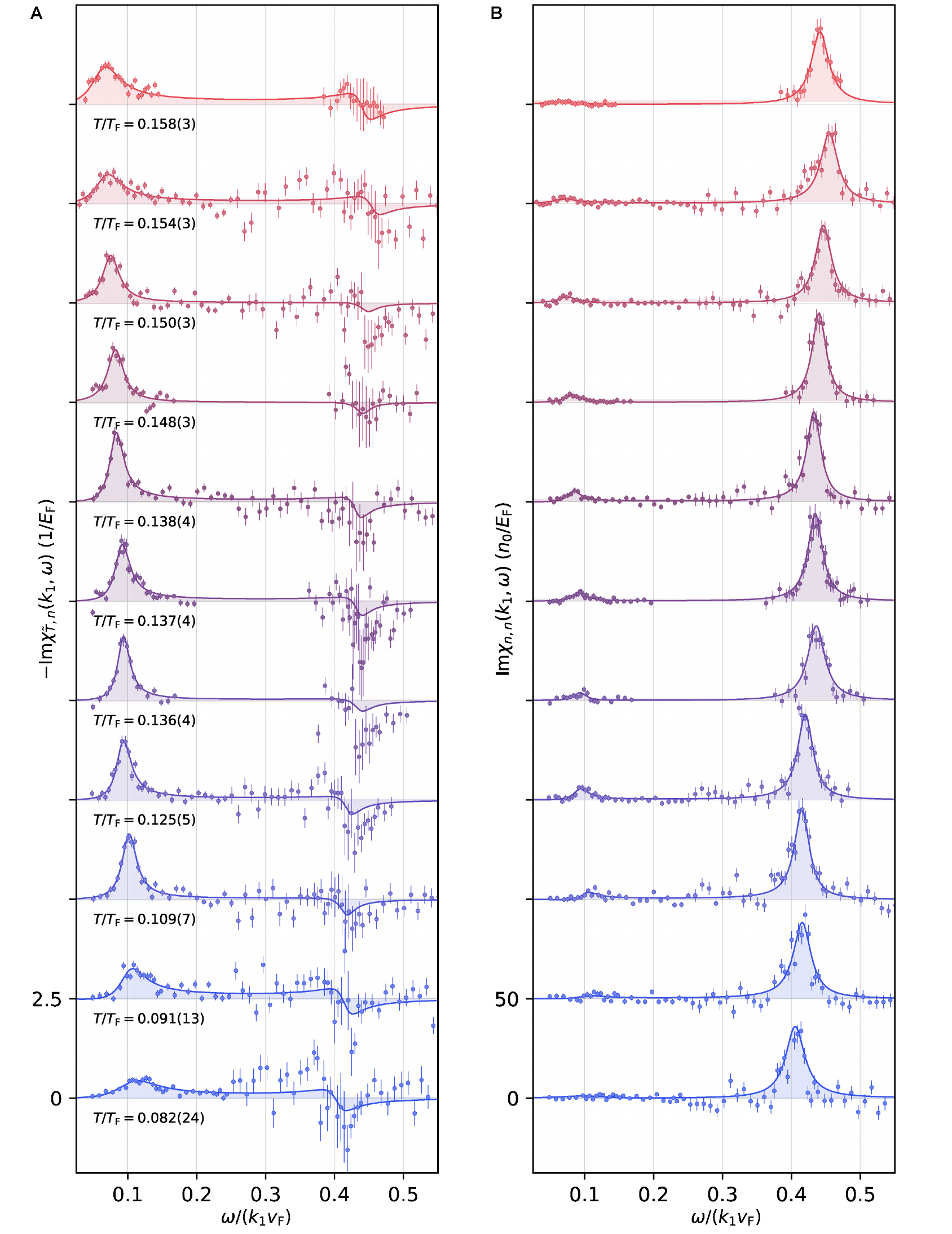}
\caption{\textbf{The normalized temperature-density response function $\mathrm{Im}\chi_{\tilde{T},n}(k_1,\omega) = \frac{\widetilde{T}}{c_V} \mathrm{Im}\chi_{s,n}(k_1,\omega)$ (A) and the density-density response function $\mathrm{Im}\chi_{n,n}(k_1,\omega)$ (B).}  The data shown here are the steady state responses from Fig.~\ref{fig:S7} normalized by the driving amplitude.
The solid lines are fits to Eq.~\ref{res:E2A} and \ref{res:E2C}.
}
\label{fig:S8}
\end{figure*}

When fitting the measured density $\Delta n(k_1,\omega)$ and normalized temperature $\Delta \tilde{T}(k_1,\omega)$ responses (Fig.~\ref{fig:S7}) with the linear response functions, we first normalize the measured responses with the corresponding modulating amplitude $U(k_1,\omega)$ to acquired the out-of-phase response functions $\mathrm{Im}\chi_{n,n}(k_1,\omega)$ and $\mathrm{Im}\chi_{n,n}(k_1,\omega)$ (Fig.~\ref{fig:S8}) using Eq.~\ref{res:E1A} and \ref{res:E1C}. A total of 8 free parameters, $c_1$, $c_2$, $D_1$, $D_2$, $\gamma$, $\frac{k_B T }{c_V} \alpha_p$, $\Gamma_{n,n}$, and $\frac{k_B T }{c_V} \Gamma_{s,n}$ are used for fitting $\mathrm{Im}\chi_{\tilde{T},n}(k_1,\omega)$ and $\mathrm{Im}\chi_{n,n}(k_1,\omega)$ simultaneously with Eq.~\ref{res:E2A} and \ref{res:E2C}.
Their number corresponds to the two resonance frequencies, the two resonance widths, and the four response amplitudes present in the response. 
The fitting curves shown in Fig.~\ref{fig:M3} and \ref{fig:S7} are products of the fitted response functions and modulating amplitude $U(k_1,\omega)$, which is a factor of $2.5$ higher at frequencies lower than $50\ \mathrm{Hz}$.
We directly obtain isentropic expansion coefficient $\gamma$ from the density-density response function (Fig.~\ref{fig:S6}A), and the value of $\frac{\kB T\alpha_p}{c_V}$ from the $\tilde{T}$-density response function (Fig.~\ref{fig:S6}B).
At unitarity, we have $c_V = \frac{3}{2} \frac{\alpha_p}{n \kappa_T}$, where $\kappa_T$ is the isothermal compressibility. Therefore, the normalized temperature-density response is governed by the compressibility of the gas: $\frac{\kB T\alpha_p}{c_V} = \frac{\kappa_T}{\kappa_0} \frac{T}{T_F}$, where $\kappa_0 = \frac{3}{2} \frac{1}{n E_F}$ is the compressibility of a non-interacting Fermi gas at zero temperature.
The measured coupling between two sound modes is weak, owed to the large difference in speeds, so that coupled and uncoupled sound speeds are almost identical, as shown in Fig.~\ref{fig:S6}C. 

While the second sound diffusivity obtained from the $m=1$ mode agrees with that from the $m=2$ mode, the width of the first sound mode for $m=1$ is inherently broadened by non-linear effects. Such non-linear behavior of the fundamental first sound mode was studied in Ref.~\cite{Zhang2021a} in the case of box-trapped Bose-Einstein condensates. Since the lowest first sound mode in the box, the sloshing mode, cannot decay into lower-lying sound modes, its damping is inherently non-linear and governed by an inverse Beliaev-type process, generating higher modes. We indeed observe the generation of $m=2$ and $m=3$ modes in the free evolution after generating the fundamental first sound mode. Accordingly, while the $m=1$ resonance width agrees with the decay rate observed for the $m=1$ mode in Ref.~\cite{Patel2020}, it is for our parameters almost twice the value one would expect from the measured $D_1$, obtained from $m=2$ in Ref.~\cite{Patel2020}. As found in Ref.~\cite{Zhang2021a} for BECs, no plateau in the damping is observed when we reduce the amplitude of excitation from $\Delta n/n_0 = 0.15$ to $0.03$. The observed non-linearity in the damping of the fundamental first sound mode will be the subject of a future investigation.

\section{Free evolution following an excitation}

\begin{figure*}
\includegraphics[width=6in]{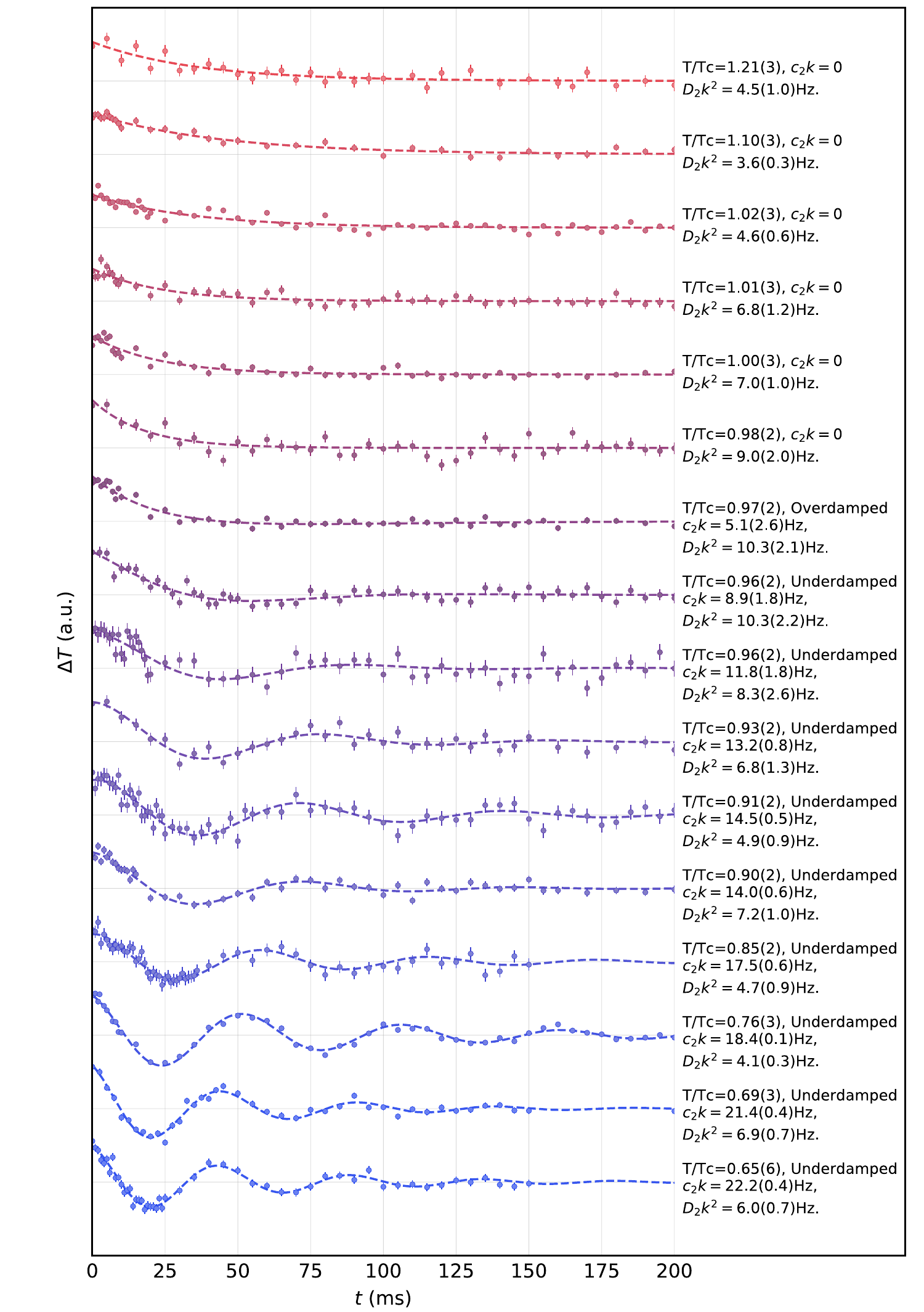}
\caption{\textbf{Free evolution of temperature changes after local heating across the superfluid transition.} The thermal amplitude $\Delta T$ is obtained at wavenumber $k_1=\pi/L$. The dashed lines are fits with Eq.~\ref{eq:evolution}, Beside each time trace, we label the fitting function used, along with the resulting second sound frequency and decay rate.
}
\label{fig:S9}
\end{figure*}

The free evolution of density, temperature, entropy etc. can also be expressed with the help of the response functions, as demonstrated by Kadanoff and Martin in Ref.~\cite{Kadanoff1963}. The general form of the time evolution of any physical quantities $A$ is
\begin{equation}
{\Delta} A (\kv,t) = \int_{-\infty}^{+\infty} \sum_{i,\pm} \left[\frac{A_{i,\pm}}{z-\omega_{i,\pm}(\kv)}  e^{-i z t}\right] \mathrm{d}z.
\label{evolution_fourier}
\end{equation}
Here $\omega_{i,\pm}(\kv)$ are the poles of the response functions at a given wave vector $\kv$. The $A_{i,\pm}$ are determined by the initial conditions and the residue of the response functions
\begin{equation}
    A_{i,\pm}=\sum_{j} \lim_{\omega \rightarrow \omega_{i,\pm}} \left [(\omega - \omega_{i,\pm})\frac{\mathrm{Im}\chi_{A,A_j}(\kv,\omega)}{\omega}  \right] \delta A_j(\kv,0),
\end{equation}
where $\delta A_j(\kv,0)$ is the initial deviation of physical observable $A_j$ from the equilibrium state and $\chi_{A,A_j}$ is the response function connecting $A$ and $A_j$.
Carrying out the inverse Fourier transform in Eq.~\ref{evolution_fourier}, one sees that the free evolution has the same form as a superposition of two damped harmonic oscillators with frequencies $c_1 k$ and $c_2 k$, and damping rates $D_1 k^2$ and $D_2 k^2$. 
The general free-evolution solution of second sound and heat diffusion is:
\begin{eqnarray}
    {\Delta} A (\kv,t)=
    \begin{cases}
    A_0 e^{-{D_2} k^2 t/2}\cos{(\sqrt{{c_{2}}^2k^2-({D_2} k^2/2)^2} t + \phi)},& c_{2}k>{D_2} k^2/2 \text{\ (a)}\\
    A_+ e^{-({D_2}k^2/2+\sqrt{({D_2}k^2/2)^2-{c_{2}}^2k^2})t}+A_- e^{-({D_2}k^2/2-\sqrt{({D_2}k^2/2)^2-{c_{2}}^2k^2})t},              & 0<c_{2}k<{D_2} k^2/2 \text{\ (b)}\\
    A_0 e^{-{D_2} k^2 t}.             & c_{2}k=0 \text{\ (c)}\\
    \end{cases}
    \label{eq:evolution}
\end{eqnarray}
The functions in Eq.~\ref{eq:evolution} are used to fit the thermal evolution data in Fig.~\ref{fig:M2}.
To determine if a time evolution trace is underdamped, overdamped, or with $c_2=0$, we first fit all time traces with both Eq.~\ref{eq:evolution}a and b to obtain fit residuals.
When the residual from fitting with Eq.~\ref{eq:evolution}a is smaller, we conclude the system is underdamped, and accept the fitted $c_2$ and $D_2$ from Eq.~\ref{eq:evolution}a.
Fitting time traces in the normal phase with $c_2=0$ using Eq.~\ref{eq:evolution}b is underconstrained, leading to a diverging fitting error.
Therefore, we categorize time traces, that show a fitting error of $c_2$ 2 times larger than the fitted value of $c_2$ with Eq.~\ref{eq:evolution}b, as purely diffusive with $c_2=0$.
We then extract $D_2$ from these data using a fit with Eq.~\ref{eq:evolution}c.
For the rest of time traces, we categorize them as overdamped and extract $c_2$ and $D_2$ using Eq.~\ref{eq:evolution}b.
A collection of measured free thermal evolution data after local heating is shown in Fig.~\ref{fig:S9}.

\section{Testing linear response of second sound}

\begin{figure*}[!ht]
\includegraphics[width=6.75in]{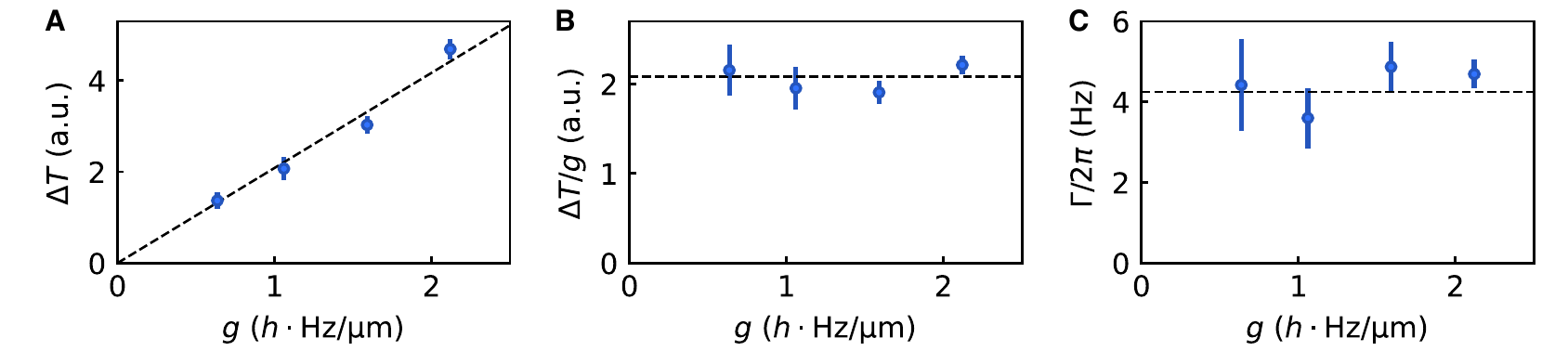}
\caption{\textbf{Testing linear response.} (\textbf{A}) The amplitude of second sound generated by resonant excitation using an oscillating potential gradient of strength $g$. The dashed line is a linear fit to the data. (\textbf{B}) Ratio between the amplitude of second sound and gradient oscillation. The horizontal dashed line is the slope of the linear fit in (A). (\textbf{C}) The damping rate of second sound measured after resonant gradient oscillation. The horizontal dashed line is the width of the reduced temperature response $\Delta\frac{T}{\TF}(k_1,\omega)$ measured at $g=2.12\ h\cdot\mathrm{Hz/\mu{m}}$. The data shown here are measured at a temperature of $T/T_\mathrm{F}=0.12.$
}
\label{fig:S10}
\end{figure*}

Proper interpretation of the obtained response functions requires the second sound excitation to remain in the linear response regime. This is verified in two distinct ways. First, we fit the thermal evolution $\Delta T (k,t)$ with the solution of a damped harmonic oscillator (Eq.~\ref{eq:evolution}) in a time range $t\geq\tau_\mathrm{f}$. By varying $\tau_\mathrm{f}$, we effectively measure the frequency and decay rate of second sound at various initial amplitudes. We find the acquired frequency $\omega=c_2 k$ and damping rate $\Gamma=D_2 k^2$ to be independent of the fitting range. Secondly, we measure the amplitude of second sound excited by a resonant gradient oscillation and find a linear dependence on the oscillation amplitude $g$ (Fig.~\ref{fig:S10}A and B). The measured damping rate is also independent of the oscillation amplitude $g$ (Fig.~\ref{fig:S10}C).

\end{document}